\documentclass[preprint,12pt]{elsarticle}
\usepackage[utf8]{inputenc} 
\usepackage[T1]{fontenc} 
   
\usepackage{xcolor}
\usepackage[normalem]{ulem}
\usepackage{amssymb}
\usepackage{algorithmic,comment}
\usepackage{array}
\usepackage{enumitem}
\usepackage{graphicx}
\usepackage{tabularx}
\usepackage{booktabs}
\usepackage{rotating}
\usepackage{multirow}
\usepackage{soul}

\usepackage[linesnumbered,ruled,vlined]{algorithm2e}

\usepackage{amsmath,amsfonts}
\usepackage{ragged2e}
\usepackage{booktabs,makecell,multirow,tabularx}
\usepackage{subcaption}
\usepackage{url}

\setcellgapes{2pt}
\newcolumntype{L}[1]{>{\RaggedRight\hspace{0pt}%
		\hsize=#1\hsize}X}

\PassOptionsToPackage{table}{xcolor}
\usepackage{tikz}
\usetikzlibrary{shapes, arrows, positioning}
\definecolor{myblue}{RGB}{135,206,250} 
\definecolor{headerColor}{RGB}{173,216,230} 
\definecolor{grayish}{RGB}{240,240,240} 
\usepackage{tabularx}
\usepackage[table]{xcolor} 
\definecolor{rblue}{RGB}{0, 122, 255}
\definecolor{contentColor}{RGB}{240,240,240} 
\definecolor{skyBlue}{RGB}{135,206,250}
\definecolor{PBblue}{rgb}{0.53, 0.81, 0.98}

\usetikzlibrary{positioning, fit, backgrounds}
\definecolor{customBlue}{HTML}{DAE8FC}
\definecolor{lightGray}{HTML}{E6E6E6}
\definecolor{fancyLabelColor}{HTML}{F8CECC}

\usepackage{bibentry}

\begin{document}
	
	\begin{frontmatter}
		
		\title{A Novel Energy-Efficient Cross-Layer Design for Scheduling and Routing in 6TiSCH Networks}
		
		\author[inst1]{Ahlam Hannachi}

		\affiliation[inst1]{organization={University of Biskra, Dept. Computer Science},
			addressline={BP 145 RP}, 
			city={Biskra},
			postcode={07000}, 
			state={Biskra},
			country={Algeria}}
		
		\author[inst2]{Wael Jaafar}
		
		\affiliation[inst2]{organization={École de Technologie Supérieure (ETS), Dept. Software and IT Engineering},
			addressline={1100 R. Notre Dame W.}, 
			city={Montreal},
			postcode={H3C 1K3}, 
			state={Quebec},
			country={Canada}}
		
		\author[inst1]{Salim Bitam}
				
		\author[inst4]{Nabil Ouazene}
				
		\affiliation[inst4]{organization={University of Batna,  Dept. Computer Science},
		addressline={Univ. Batna}, 
		city={Batna},
		postcode={05000}, 
		state={},
		country={Algeria}}

		\begin{abstract}

            {}{The 6TiSCH protocol stack plays a vital role in enabling reliable and energy-efficient communications for the Industrial Internet of Things (IIoT). However, it faces challenges, including prolonged network formation, inefficient parent switching, high control packet overhead, and suboptimal resource utilization. To tackle these issues, we propose in this paper a novel cross-layer optimization framework aiming to enhance the coordination between the Scheduling Function (SF), the Routing Protocol for Low-Power and Lossy Networks (RPL), and queue management. Our solution introduces a slot-aware parent switching mechanism, early slot reservation to mitigate queue overflow, and a refined slot locking strategy to improve slot availability. To reduce control overhead, the proposed method merges 6P cell reservation information into RPL control packets (DIO/DAO), thus minimizing control exchanges during parent switching and node joining. Optimized slot selection further reduces latency and jitter. Through extensive simulations on the 6TiSCH simulator and under varying network densities and traffic loads, we demonstrate significant improvements over the standard 6TiSCH benchmark in terms of traffic load, joining time, latency, and energy efficiency. These enhancements make the proposed solution suitable for time-sensitive IIoT applications.}
		\end{abstract}
		
		
				
		
	\end{frontmatter}
	
    \section{Introduction}
    \label{sec:sample:appendix}
    {}{The Industrial Internet of Things (IIoT) is reshaping critical sectors such as industrial automation, smart grids, and environmental monitoring by enabling large-scale, real-time data exchange across vast sensor networks. However, achieving ultra-reliable, low-latency, and energy-efficient communication in such resource-constrained environments remains a major challenge. Current networking protocols struggle to provide deterministic performance while balancing energy efficiency, scalability, and real-time constraints, making protocol enhancements crucial for IIoT applications. To meet these stringent requirements, 6TiSCH has emerged as a key protocol stack, integrating IPv6 over the Time-Slotted Channel Hopping (TSCH) mode of IEEE 802.15.4e. By leveraging time-slotted access, multi-channel communication, and channel hopping, 6TiSCH ensures deterministic reliability, enhances interference mitigation, and provides efficient energy management, making it a strong candidate for time-sensitive IIoT applications.}
           
	The 6TiSCH protocol encapsulates IPv6 over the TSCH mode of IEEE 802.15.4e, marking a significant advancement in networking infrastructure for IIoT. This protocol merges the precise and predictable TSCH mechanism with the broad capabilities of IPv6, thus providing industrial-level reliability and precise timing. Indeed, 6TiSCH is designed for low-power and lossy networks (LLNs), which are typical in IIoT environments due to their limited power and unreliable transmission media. This makes it particularly effective for seamless integration and interoperability among low-power IoT devices across various environments. 	

{}{6TiSCH ensures reliable, low-latency communication, making it particularly suitable for time-sensitive applications such as industrial automation, healthcare, and environmental monitoring. Its structured scheduling mechanism supports real-time data transmission and synchronization, which are critical for these domains. However, existing scheduling approaches lack adaptability, leading to inefficient slot utilization and degraded real-time performance in IIoT deployments.} 
{}{Indeed, for real-time IIoT applications, minimizing jitter and latency is essential to ensure stable packet transmission and synchronization. However, inefficiencies in 6TiSCH’s scheduling and queue management mechanisms introduce unpredictable delays, negatively affecting control loops and time-sensitive decision-making processes.}
{}{Moreover,} energy efficiency in IIoT stands at the forefront of technological challenges, with researchers globally striving to meld enhanced performances with reduced energy consumption. Specifically, the medium access control (MAC) layer is pivotal in controlling the energy drain 
    by regulating access to the shared communication medium, typically a wireless channel. The TSCH protocol stands out as the leading MAC operational mechanism and is included in several standards, such as ANSI/ISA100.11a and WirelessHart IEC62591 {\cite{inproceedings2018}, \cite{9016059}}. 

	\begin{figure}
	\centering
	\includegraphics[width=0.8\linewidth]{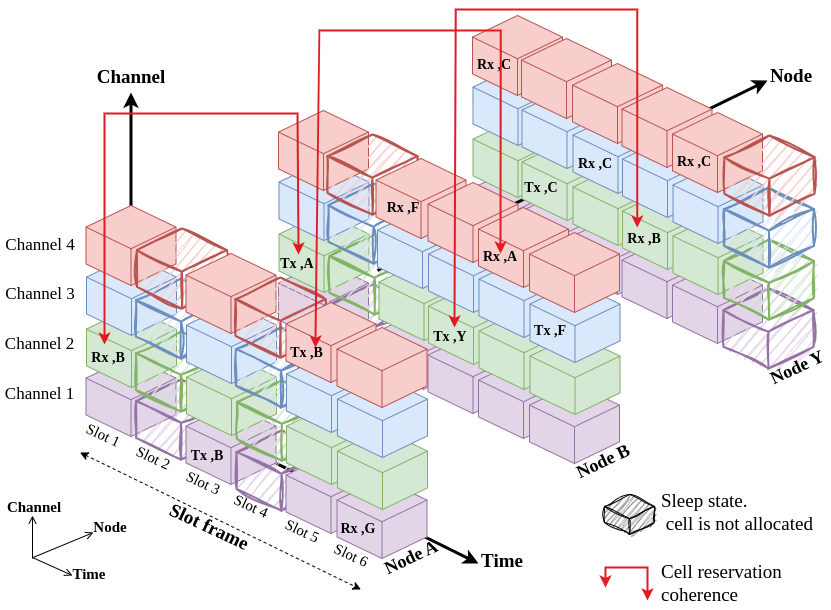}
	\caption{TSCH schedule of nodes A, B, and Y.}
	\label{fig:tsch_all}
\end{figure}

The core concept of TSCH lies in integrating Time Division Multiple Access (TDMA) with channel-hopping techniques. This fusion results in a dual medium access mechanism based on time and channel. In this context, ``Time'' is referred to by \textit{time slot}, and ``Frequency'' is referred to by \textit{channel}. A time slot within a specific channel is defined as a \textit{cell}. For every time slot, shown in Fig. \ref{fig:tsch_all}, an IoT node specifies its activity in a given channel, choosing to transmit to a node (Tx), receive from a node (Rx), or sleep, with the latter not requiring any specific cell selection \cite{7460875}. {}{Moreover, a ``slotframe'' represents a repeating schedule of time slots.}
The efficiency of the TSCH schedule is reflected in the consistent actions of devices. Thus, each device has to judiciously select the combination of (state, cell, neighbor), where the \textit{state} $\in \left\{Tx, Rx, Sleep \right\}$. For instance, in node $A$'s TSCH schedule in Fig. \ref{fig:tsch_all}, the cell \textit{(slot 1, channel 2)} is set for receiving from node $B$. Conversely, in node $B$'s schedule, the same cell is allocated for transmitting to node $A$.
	Moreover, $A$ is set to sleep mode during time slots $2$ and $4$, while $B$ enters sleep mode in time slot $2$.

	

\begin{figure}
	\centering
	\includegraphics[trim={0.3cm 4cm 2.5cm 0.1cm},clip,width=0.8\linewidth]{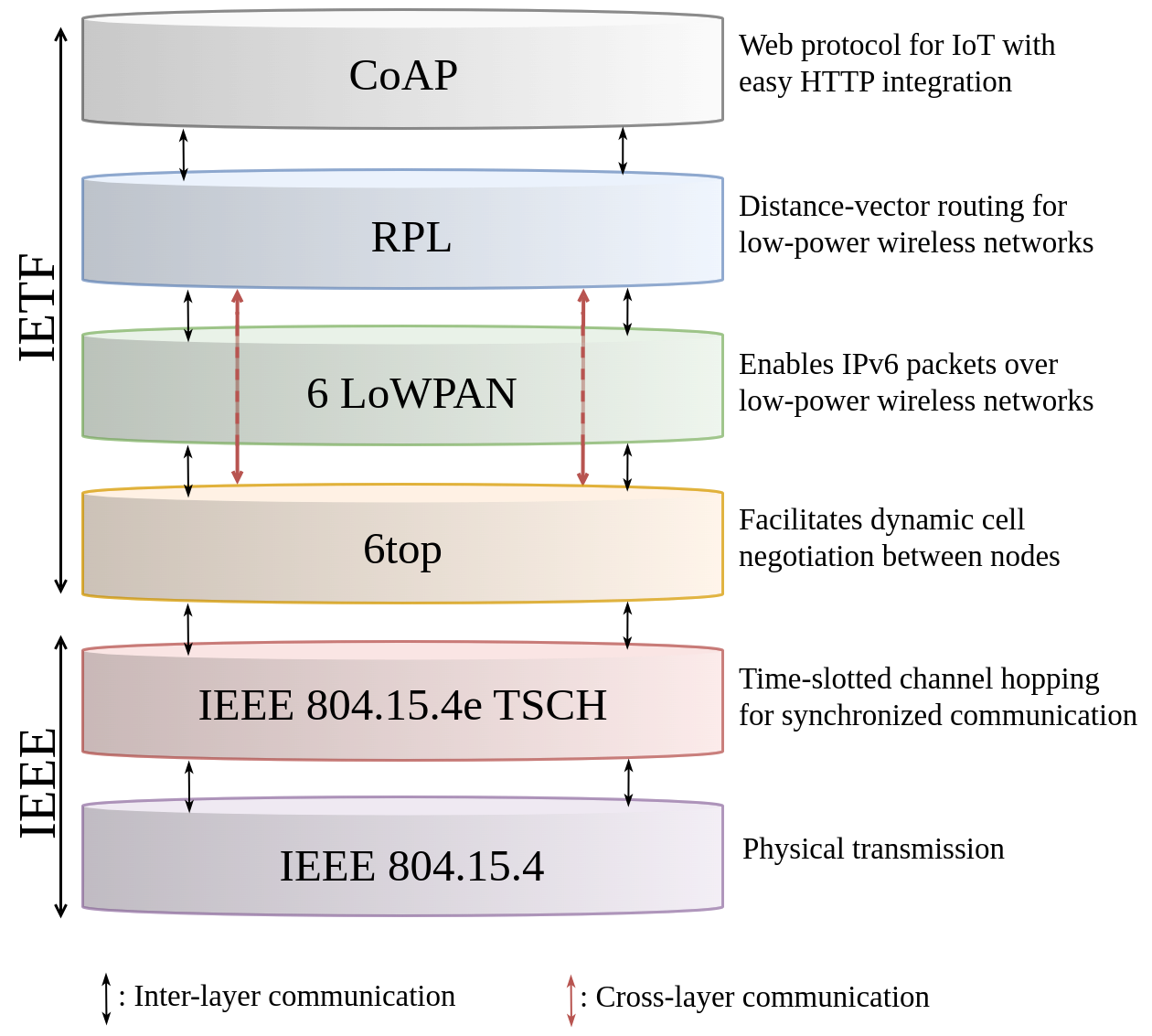} 
	\caption{The 6TiSCH protocol stack \cite{tabouche_traffic-aware_2023}.}
	\label{fig:6tisch_layers}
\end{figure}

{}{Building on TSCH, the Internet Engineering Task Force (IETF) established the 6TiSCH working group to define protocols that seamlessly integrate IPv6 over TSCH, enabling efficient network formation, routing, and resource scheduling in low-power industrial environments.} The 6TiSCH protocol stack is depicted in Fig.\ref{fig:6tisch_layers}. Specifically, the 6top layer holds the Scheduling Function (SF) responsible for TSCH cell allocation and traffic demands/performance metrics monitoring. The conventional SF is the Minimal Scheduling Function (MSF) \cite{rfc9033}.	
	Moreover, the Routing Protocol for Low-Power and Lossy Networks (RPL) layer, which encompasses the Objective Function (OF), establishes and sustains the routing topology. It does so by constructing a Destination-Oriented Directed Acyclic Graph (DODAG), a structured network topology that enables optimized routing. Through DODAG, RPL efficiently propagates routing information, tailoring pathways based on various critical metrics to meet the specific demands of the network. Also, RPL allows IoT nodes, in particular Industrial IoT (IIoT), to discover their best potential parents and establish a route towards the network root. The conventional OF of RPL is the Objective Function Zero (OF0) \cite{rfc6552}. 
	
	Both RPL and 6top utilize control packets to maintain and optimize the IoT network's operation, namely DODAG Information Solicitation (DIS), DODAG Information Object (DIO), and Destination Advertisement Object (DAO) for RPL, and 6P transactions in 6top. The relation between RPL and 6top is reciprocal. For instance, while RPL alters the routing topology by changing a node’s parent, 6top dynamically adjusts the scheduled cells to accommodate this change, i.e., releasing assigned cells to the old parent and negotiating new cell allocation with the new parent. Moreover, RPL determines the next-hop parent for data transmission based on its topology, while 6top ensures cell availability for this specific transmission according to the TSCH schedule.  However, this cross-layer interaction between the RPL and 6top layers (shown by red lines in Fig.\ref{fig:6tisch_layers}) implicitly controls very critical performance metrics, such as the joining time of a new node to the DODAG network and energy efficiency.  \\

    {}{At the intersection of SF and RPL, inefficiencies in routing and scheduling coordination result in wasted network resources, increased energy consumption, and higher packet loss. A major contributor to this problem is the inefficient parent switching mechanism. In 6TiSCH, nodes may switch to a new parent without verifying slot availability, leading to network instability and inefficient resource allocation. Another challenge arises when a node joins the network but experiences significant delays before becoming fully operational. Upon joining the DODAG, a node must undergo multiple control packet exchanges before it can transmit data. Since these packets are sent in separate time slots, the process of becoming operational is delayed, preventing the node from contributing to network traffic immediately. This inefficiency is exacerbated in dense networks, where multiple joining nodes compete for minimal shared cells, further extending the delay. Beyond SF-RPL coordination challenges, inefficiencies in the coordination between queue management and slot allocation can degrade network performance. A major issue is queue overflow, which occurs when a node accumulates packets but lacks available transmission cells, leading to packet drop. Specifically, when the transmission cells to the destination are lacking, the queue fills up, causing packet drops before transmission.
    Finally, a critical inefficiency stems from slot reservation delays caused by cell locking. Specifically, when a node proposes a set of cells for allocation, the latter are locked until either receiving a response or timing out. During this period, the slots remain unavailable for reservation, leading to resource losses and delayed slot reassignment.}

    {}{To address these challenges, we propose a novel cross-layer optimization framework designed to enhance 6TiSCH network performances.}
    To the best of our knowledge, very few works have investigated the cross-layer design of scheduling (6top) and routing (RPL) functions within 6TiSCH networks and their generated advantages and limitations. {}{Hence, the main contributions of this work can be summarized as follows:}

    \begin{enumerate}
        \item {}{We propose a novel slot-aware parent switching mechanism where nodes check slot availability before switching, ensuring more stable routing decisions and reducing unnecessary reconfigurations.}

        \item {}{We introduce parallel control packet transmissions to reduce the node activation time.}

        \item {}{To attenuate packet drops, we develop an early slot reservation strategy where nodes dynamically request additional cells before queue overflow occurs.}

        \item {}{To improve slot availability and resource usage, we propose a flexible slot locking mechanism that limits the duration for which proposed slots remain reserved without confirmation.}
        \item {}{Through extensive simulations in the 6TiSCH simulator \cite{6TiSCH_Tutorial}, we demonstrate the efficacy of our framework in improving the joining time, energy efficiency, packet delivery, and jitter performance, making it a robust enhancement for IIoT.}
    \end{enumerate}

{}{The remainder of this paper is structured as follows: Section 2 provides background on 6TiSCH and reviews related works. Section 3 identifies key challenges and inefficiencies in 6TiSCH networks. Section 4 details the proposed solution, explaining its design and implementation. Section 5 evaluates the performance of our approach through simulations, comparing it to the MSF benchmark. Section 6 discusses the challenges and future research directions. Finally, Section 7 concludes the paper.}

	
	\section{6TiSCH Background and Related Work{}{s}}
	
	In the field of 6TiSCH networks, ongoing research is dedicated to developing effective, reliable, and energy-efficient communication strategies. This often involves examining and refining existing standards. In this section, we first outline the key conventional protocols within the 6TiSCH stack.
	Then, we highlight recent related works. 
	
	\subsection{6TiSCH Background}
	
	The 6TiSCH standard lays the groundwork for the operation of IoT networks. We present their main components relevant to our research, namely the Minimal Configuration (MC) and the Minimal Scheduling Function (MSF). The MC details the procedure for the node joining process \cite{rfc8180}, while the MSF defines the scheduling mechanism, which outlines how communication cells are assigned and released \cite{ietf-6tisch-msf-10}.
	\subsubsection{Minimal Configuration (MC)}
	The MC defines the sequential steps that each node must undertake before becoming operative and start transmitting data. Its steps, shown in Fig. \ref{fig:network_formation}, are described as follows: 
	
	\begin{figure}
		\centering
		\includegraphics[trim={0.1cm 2cm 1.5cm 1cm},clip,width=0.65\linewidth]{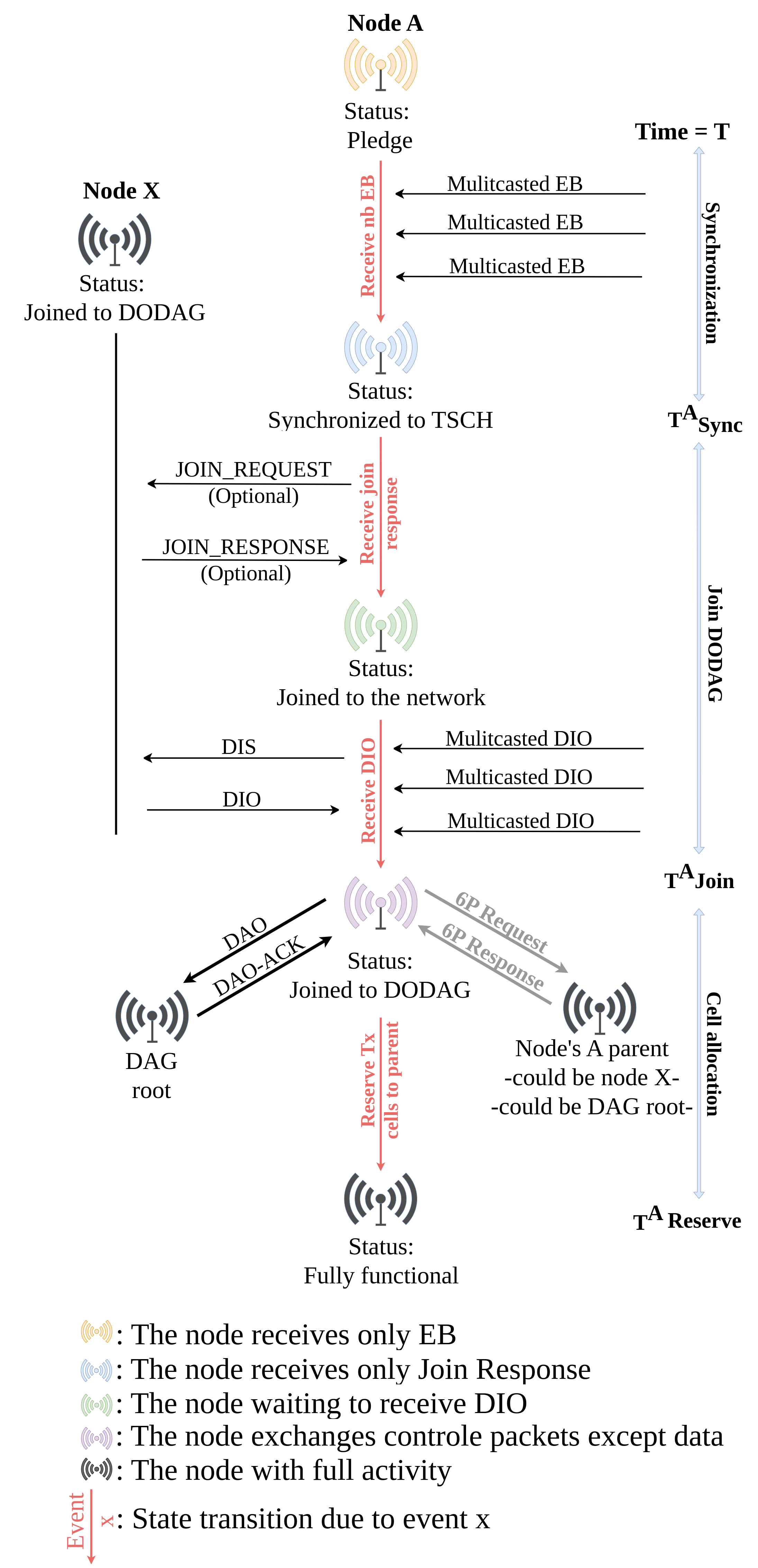}
		\caption{6TiSCH network formation using MC.}
		\label{fig:network_formation}
	\end{figure}
	
	\begin{itemize}
		\item \textit{Synchronization:} Initially, each node is in a ``pledge'' state, passively listening on randomly selected channels to anticipate Enhanced Beacons (EBs) sent by nodes that are already part of the network, including the \textit{root node} that initiates the network formation. Upon the reception of a predefined number of EB packets, a ``pledge'' synchronizes with the IoT network. For instance, node A in Fig. \ref{fig:network_formation} starts in the ``pledge'' status at time $T$ and synchronizes at time $T^A _{\rm Sync}$, where energy is expanded during the interval $T_1 =T^A _{\rm Sync}-T$.
		
		\item \textit{DODAG joining:} After synchronization, if secure joining is required, the node sends a ``JOIN REQUEST'' to the EB source and awaits the ``JOIN RESPONSE''. Otherwise, it directly sends a DODAG Information Solicitation (DIS) packet. The DIS mechanism operates in two distinct modes:
		\begin{itemize}
			\item \textit{Unicast DIS:} In this mode, the receipt of a DIS message prompts the targeted node to respond with a DIO message.
			\item \textit{Broadcast DIS:} In this mode, receiving a DIS message leads nodes to reset their DIO Trickle timers, thereby influencing the timing of subsequent DIO message dissemination.
		\end{itemize}				
		The DODAG joining is driven by the periodic multicasting of DIO messages by nodes that have previously integrated the DODAG. DIO messages encapsulate vital information regarding the DODAG, including its identification (ID), version, and the node's rank within the structure. DIO packets are delineated into:
		\begin{itemize}
			\item \textit{Periodic DIO:} 
			These messages are needed to establish and sustain the topology of a DODAG. They provide critical information about the DODAG's configuration, including the sender's rank and set of nodes utilized for parent selection within the DODAG. 
			The period of the DIO transmission is defined by the Trickle algorithm, and according to the network's consistency \cite{rfc6206}.
			\item \textit{DIO as a response:} It is a unicast response to a DIS packet sent by a soliciting node asking for relevant routing information. For instance, node A in Fig. \ref{fig:network_formation} joins the DODAG at time $T^A_{\rm Join}$, where the time interval $T_2 = T^A_{\rm Join}- T^A_{\rm Sync}$ corresponds to the joining procedure.
		\end{itemize}
		
		\item \textit{Cell allocation:} Once synchronized, every node reserves its first cell, referred to as the ``minimal cell'', at time slot $0$ and channel $0$. Following this initial reservation, when a node receives a DIO message, it selects a parent among neighbors that advertised the DIO messages, according to the RPL's objective function. Then, the node triggers the process of cell reservation with the new parent by initiating a 6P transaction and transmits a DAO packet to the DAG root to indicate its parent, so that the DAG root can build/update the topology tree. If the parent is the DAG root itself, then the node sends both the 6P transaction and DAO packet to it. DAO packets are destined to the DAG root to facilitate the update of the routing tree. They come in two types:
		\begin{itemize}
			\item \textit{Periodic DAO}: It is transmitted periodically and used to maintain and refresh the route information at the DAG root.
			\item \textit{Parent Switching DAO}: It is sent immediately after a parent-switching event to swiftly notify the DAG root of changes in the routing topology. 
		\end{itemize}
	\end{itemize}

\begin{figure}
		\centering
		\includegraphics[width=0.8\linewidth]{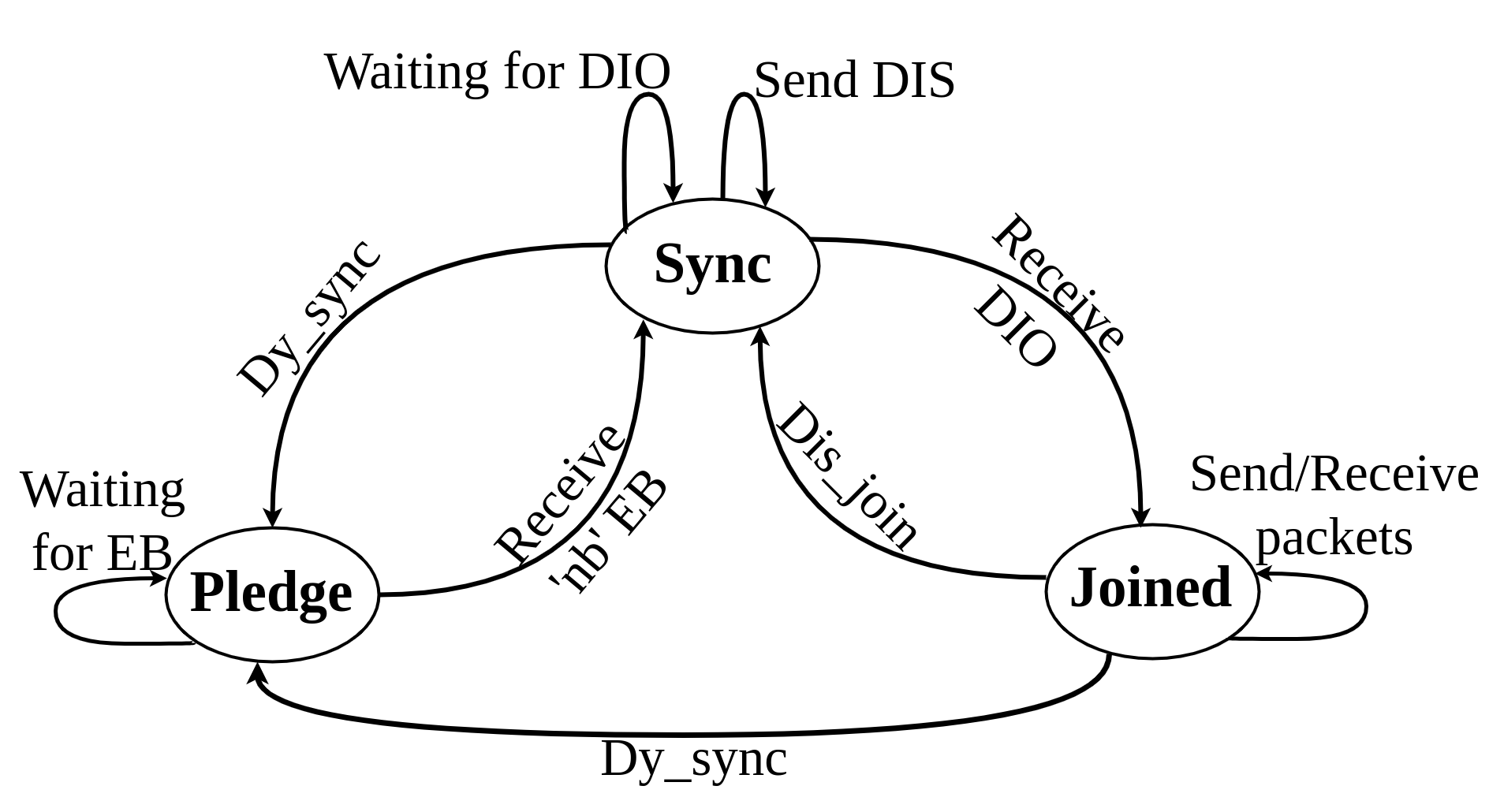}
		\caption{Node status cycle.}
		\label{fig:status_cycle}
	\end{figure}

	In summary, nodes within 6TiSCH networks dynamically alter their operational states throughout their life cycle, influenced by localized parameters and network conditions. These transitions and state changes, illustrated in Fig. \ref{fig:status_cycle}, embody the node's continual attempt to preserve network connectivity, synchronization, and data transmission integrity within the inherently challenging communication environment of 6TiSCH networks. Notably, a node reverts to its initial ``pledge'' state if it fails to receive any packet within a predefined temporal window (\textit{Dy\_sync}), indicative of potential communication disruptions or isolation from the network. Furthermore, if the node experiences a linkage failure to its parent or while executing parent switching (\textit{Dis\_join}), it regresses to a ``synchronized'' state.
	
	\subsubsection{Scheduling Function (SF)}
\begin{figure}
		\centering
		\begin{subfigure}{0.7\linewidth}
			\includegraphics[width=\linewidth]{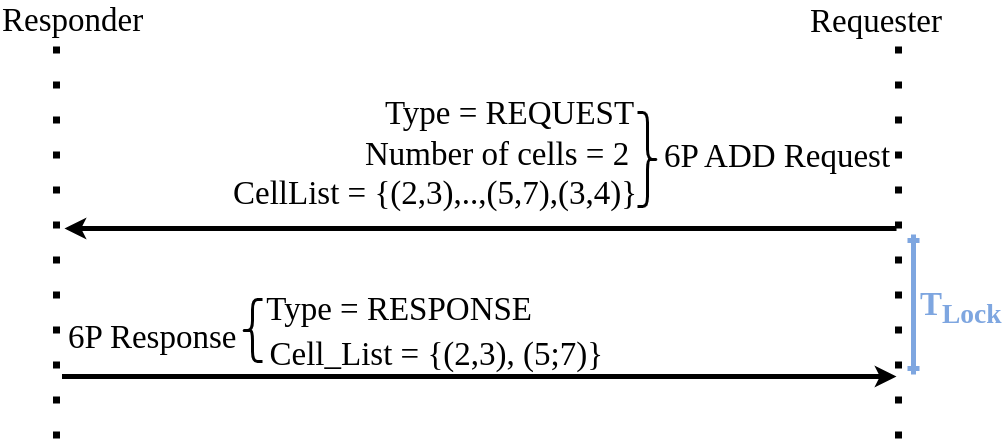}
			\caption{2-Step 6P transaction}
			\label{fig:6p-2step}	
		\end{subfigure}
		\hspace{0.1cm}
		\begin{subfigure}{0.7\linewidth}
			\includegraphics[width=\linewidth]{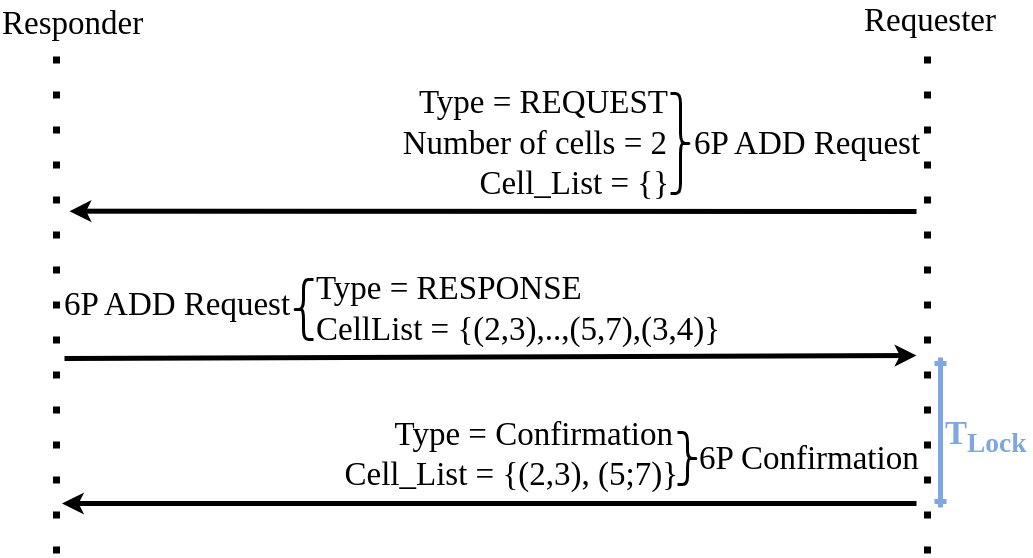}
			\caption{3-Step 6P transaction.}
			\label{fig:6p-3step}	
		\end{subfigure}
		\caption{Examples of 6P transactions.}
		\label{fig:6p}	
	\end{figure}

	The scheduling function, SF, serves as a critical mechanism to allocate time slots and channels (i.e., cells) within the TSCH mode. Its primary role is to manage the scheduling of communications across the time-frequency domain. By dynamically responding to the varying demands and conditions of the network, SF ensures the judicious use of resources and maintains dependable communications. As shown in Fig. \ref{fig:6p}, the SF conducts the cell reservation procedure using 6P protocol transactions, where the term ``Requester'' denotes a node in need of reserving new cells from another node, referred to as the ``Responder''. In this process, the ``Requester'' initiates the cell reservation process \cite{rfc8480,ietf-6tisch-msf-10}. There are two distinct approaches: the 2-step and 3-step methods. The key difference lies in which node proposes the list of cells. In the 2-step transaction illustrated in Fig. \ref{fig:6p-2step}, the ``Requester'' proposes the list of cells, which leads to exchanging two control packets, whereas in the 3-step transaction, presented in Fig. \ref{fig:6p-3step}, the ``Responder'' proposes the list of cells, thus exchanging three control packets. Nevertheless, in both approaches, the proposed cells are locked during the period $T_{\rm Lock}$.
	The conventional SF of the 6TiSCH protocol stack is the Minimal Scheduling Function (MSF), which follows a strategy of random cell selection, both in cell proposal during 6P transactions and in cell reservation among the proposed/available cells. Nevertheless, the nodes dynamically determine the number of cells needed based on their traffic requirements and network conditions. Although this stochastic approach provides adaptability to varying traffic conditions, it may face challenges such as internal collisions, particularly in dense networks. This raises concerns about its applicability in scenarios with stringent Quality-of-Service (QoS) requirements.
	
    \subsection{Related Works}
	
	The conventional 6TiSCH mechanisms, while foundational, have been scrutinized for potential limitations and areas of enhancement. This section investigates the existing literature for proposed methodologies, achievements, and gaps, thereby contextualizing the motivation and novelty of our work. 
	
	Aiming to enhance fundamental aspects of 6TiSCH, particularly the TSCH configuration, {}{the} authors of \cite{PMID:33668935} adopted {}{a} heterogeneous time slot length 
    to minimize communication latency. {}{However, the use of multiple minimal cells for multi-PHY discovery, dynamic slot duration adaptation, and the added control overhead from frequent slot reconfiguration increase network overhead, memory usage, scheduling complexity, and energy consumption. In} \cite{ParentPhy}, the authors introduced a heuristic method for parent selection in multi-PHY TSCH, showing that slot bonding can adapt to different PHY data rates and achieve high Packet Delivery Ratios (PDRs). 
    {}{Regarding 6TiSCH network formation, the study in \cite{8502509} evaluated the network formation procedure and identified several limitations.} Then, the authors proffered several recommendations and established guidelines for parameter adjustments, tailoring parameters to several network topologies. Also, the authors of \cite{kalita_opportunistic_2021} addressed performance degradation during 6TiSCH network formation. They highlighted the challenges faced by new nodes joining the network, primarily due to the policy of allocating only one shared cell per slotframe for control packet transmission. They particularly emphasized the limitations of prioritizing EBs during network formation. Then, they proposed an improved approach that effectively integrates control packets into the network formation process. {}{In \cite{MOHAMADI2021103026}, the authors aimed to accelerate the network formation process through the use of active scanning procedures and a Trickled beacon advertising strategy. In their approach, nodes attempting to join transmit beacon request packets, thereby increasing the likelihood of detection by already associated nodes and reducing the overall association time. Although this solution significantly improved network formation, it overlooked energy efficiency as the joining nodes remain in continuous listening mode during the formation phase.} {}{In \cite{Park2025}, the authors proposed the Enhanced Network Bootstrapping technique using Multi-Channel (EBMC) to improve the network formation. Since EBMC relies on a static channel selection algorithm based on the nodes' EUI-64 addresses, it suffers from persistent synchronization failures when constant interference affects the assigned channel, thus preventing nodes from joining the network. Furthermore, since topology formation in EBMC does not incorporate dynamic parent selection, inefficient multi-hop routes may occur, resulting in higher end-to-end latency and routing overhead. Despite that multi-channel allocation reduced collision probability, EBMC’s static control packet assignments led to underutilized channel resources during periods of low control packet activity.} 
    
    {}{Researchers also explored optimizing scheduling functions to enhance network efficiency.} Typically, research works that enhance SF fall into three categories: 1) Centralized approaches \cite{ysf,ReSF,FafoutisEOPC18}, 2) distributed approaches \cite{article_kim_2021}, and 3) autonomous approaches \cite{TESLA}. In \cite{ysf}, the authors proposed a fully centralized schedule for data gathering, where each node has to send the allocation request to the DAG root and receive the answer, thus wasting a considerable amount of energy and time. Alternatively, in \cite{ReSF}, a central entity is responsible for managing the scheduling of transmissions in the network and uses a recurrent neural network (RNN)-based model to predict the next transmission slot for each node, aiming to reduce the latency of end-to-end communications. However, a higher control packet overhead is generated. {}{Thus, it imposes a huge burden on the nearest node to the root, and the network formation process takes a long time and consumes a considerable energy, particularly in dense networks.} 
    {}{Alternatively}, the authors of \cite{FafoutisEOPC18} introduced an Adaptive Static Scheduling (ASS) technique designed to enhance the network's energy efficiency. {}{In \cite{article_kim_2021}, a distributed mechanism to improve routing and scheduling was introduced. However, the allocation of dedicated control cells to minimize collisions led to inefficient resource utilization, particularly in dense and highly mobile deployments. Moreover, the reliance on fixed downlink routes, while beneficial for latency reduction, increased complexity in dynamic environments due to frequent updates and higher control overhead. Finally, the frequent exchange of control messages, route updates, and dedicated cell allocations degraded energy efficiency. In contrast, \cite{TESLA} proposed TESLA, a traffic-aware elastic slotframe adjustment scheme that allows each node in the network to dynamically adjust its slotframe size based on real-time traffic conditions.} 
    
    
	Other works focused on optimizing the RPL's objective function. For instance, authors of \cite{inproceedings} introduced a novel parent selection function that accounts for the number of neighbors and traffic load. It optimized the PDR to minimize frequent parent switching. In \cite{inproceedingsBekar}, a Q-learning-based algorithm is proposed to refine the RPL's objective function, while \cite{10225947} proposed an early parent switching scheme based on the remaining queue capacity of the parent node. 
    {}{However, its reliance on frequent parent switching and queue monitoring introduces control overhead. Also, it does not account for the fact that a parent’s queue congestion may stem from inefficiencies in the SF rather than true traffic imbalance, potentially leading to suboptimal routing decisions. In addition, \cite{inproceedingsBekar} suffers from high computational complexity as it uses Q-Learning, which requires frequent Q-value updates, large table storage, and intensive processing, making it less suitable for resource-constrained 6TiSCH networks.} 
	
    {}{To address limitations inherent to single-layer solutions, cross-layer approaches have been proposed to jointly optimize multi-layer parameters. For instance, authors in \cite{AIMARETTO2023100988} proposed a cross-layer scheduling and routing approach that enhances reliability in 6TiSCH networks by integrating alternative parent selection, MPLS-like source routing, and Bounded Delay Packet Control (BDPC). By leveraging multiple alternative parents and implementing a deadline-aware packet forwarding mechanism, their method significantly improved the packet delivery ratio and ensured bounded latency. However, this comes at the cost of higher computational complexity as each node has to process per-packet label switching, maintain multiple parent paths, and track delivery deadlines dynamically. These additional demands increase overhead and are not well-suited for resource-constrained IIoT devices. Also, the approach proposed in \cite{article_Qlearn_Fawaz} improved packet delivery and latency but introduced high computational complexity, increased communication overhead for Q-table sharing, and required careful parameter tuning.
     In \cite{inproceedingsIMSF}, the authors proposed the Improved Minimal Scheduling Function (IMSF) to enhance MSF by dynamically adjusting the number of allocated cells based on queue occupancy and link quality (ETX). This technique improved adaptability to traffic variations and reduced packet loss without generating additional control packets beyond standard 6P transactions. However, IMSF’s dependence on ETX metrics and historical queue data may limit its responsiveness in dynamic network conditions.}
 
 {}{As discussed above, existing solutions either introduce additional control overhead, rely on complex computations, or do not adapt to dynamic network conditions, often leading to high delays and energy consumption. Moreover, most works enhanced a single performance metric at the cost of others.
    Consequently, in this paper, we propose a novel cross-layer approach that accelerates network formation, optimizes scheduling, and improves adaptability while minimizing control overhead, latency, and energy consumption, making it a lightweight yet efficient solution for energy-constrained IIoT environments.}
 
    \section{Problem Identification}
		
    Several operational challenges impede the effectiveness of deploying the 6TiSCH protocol within IIoT networks. They include:

       \begin{figure}
			\centering
			\includegraphics[width=0.7\columnwidth]{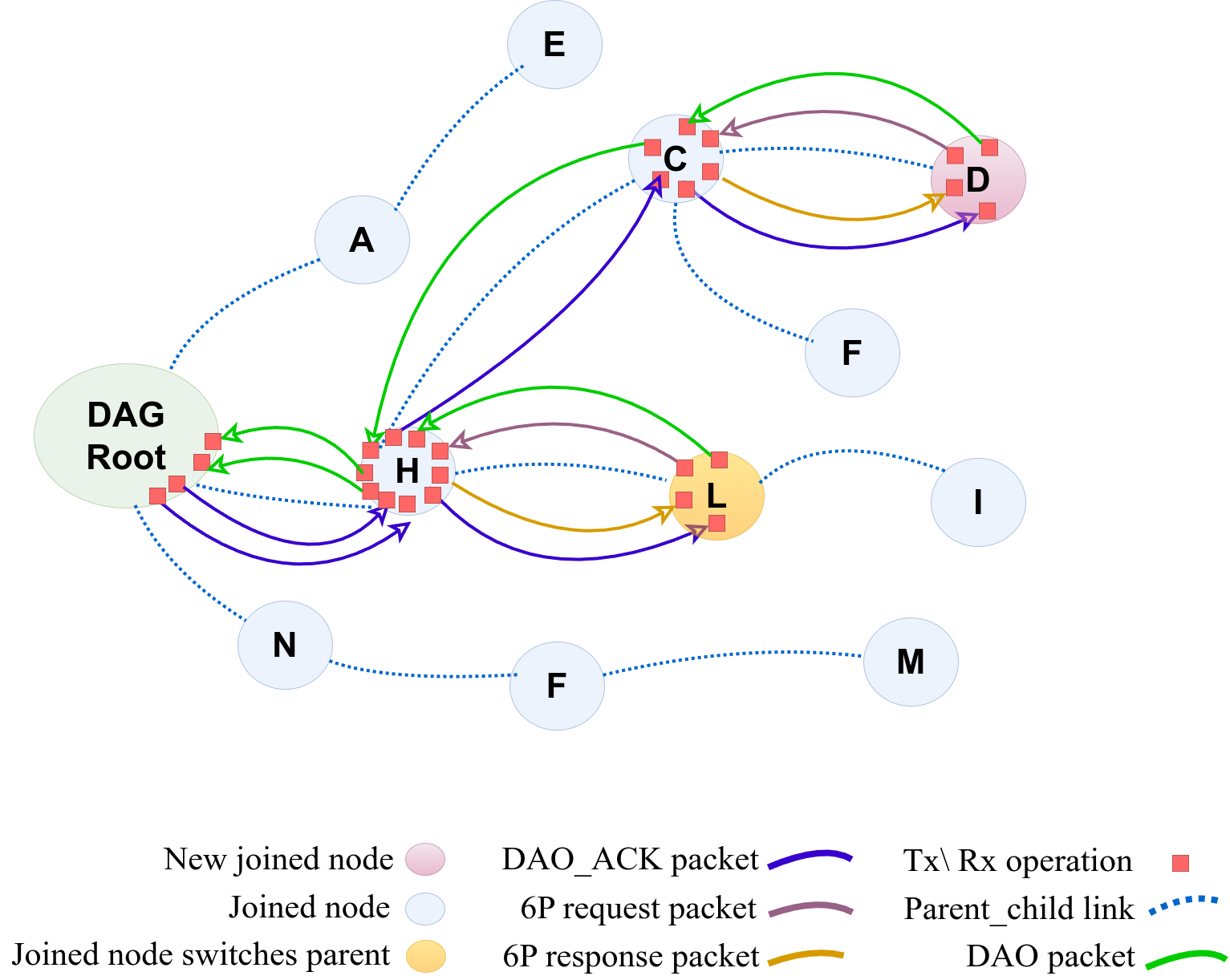}
			\caption{Control packets triggered upon parent switching and joining DODAG \cite{10811247}.}
			\label{fig:ctrl_pkt}
		\end{figure}

    \begin{enumerate}
			\item \textit{Complex Time Slot Allocation for External Nodes:}	
			The 6TiSCH protocol 
            {}{experiences difficulties to timely allocate} sufficient time slots for nodes not directly part of the DODAG. This issue is critical in scenarios where nodes keep moving, {}{and continuously changing the DODAG's topology, thus requiring quick slot allocation to maintain connectivity.} 
            {}{As depicted in Fig. \ref{fig:ctrl_pkt}, newly joined nodes, such as $D$, send DAO packets to the DAG root, initiate 6P requests to their parents, and await corresponding 6P responses, before effectively starting to transmit data. However, if a joining node repeatedly fails to send DAO or 6P messages due to unsuccessful transmission or lack of response from the parent, it will eventually lose synchronization and revert to a desynchronized state (See Fig. \ref{fig:status_cycle})}. A lack of timely integration into the network delays operational synchronization and increases energy consumption as new nodes repeatedly attempt to join the network.

	
			\item \textit{Excessive Use of Control Packets {}{(6P)}:}	
			The current implementations of the 6TiSCH scheduling function (SF) and routing protocol (RPL) rely heavily on control packets, thus leading to a congested network, which is particularly problematic in large-scale systems 
            {}{A large control packets traffic increases competition for transmission opportunities, which can cause queuing delays, higher latency, and degraded system reliability. This problem is critical in dense networks where simultaneous control packet exchanges occur frequently during network formation, parent switching, and node joining.} {}{Fig. \ref{fig:ctrl_pkt} illustrates the issue where nodes $L$ and $D$ experience frequent packets exchange during parent switching and DODAG joining, respectively. 
            Both events trigger DAO packets that traverse multiple intermediate nodes before reaching the DAG root, which responds with DAO-ACK messages. Concurrently, the $L$ and $D$ initiate 6P requests to negotiate cell reservations with their new parents, who reply with 6P responses. Such simultaneous DAO and 6P exchanges overload intermediate nodes, create bottlenecks near the DAG root, and delay data delivery. 
            }


    \begin{figure}
        \begin{subfigure}{0.5\linewidth}
            \includegraphics[width=0.98\linewidth]{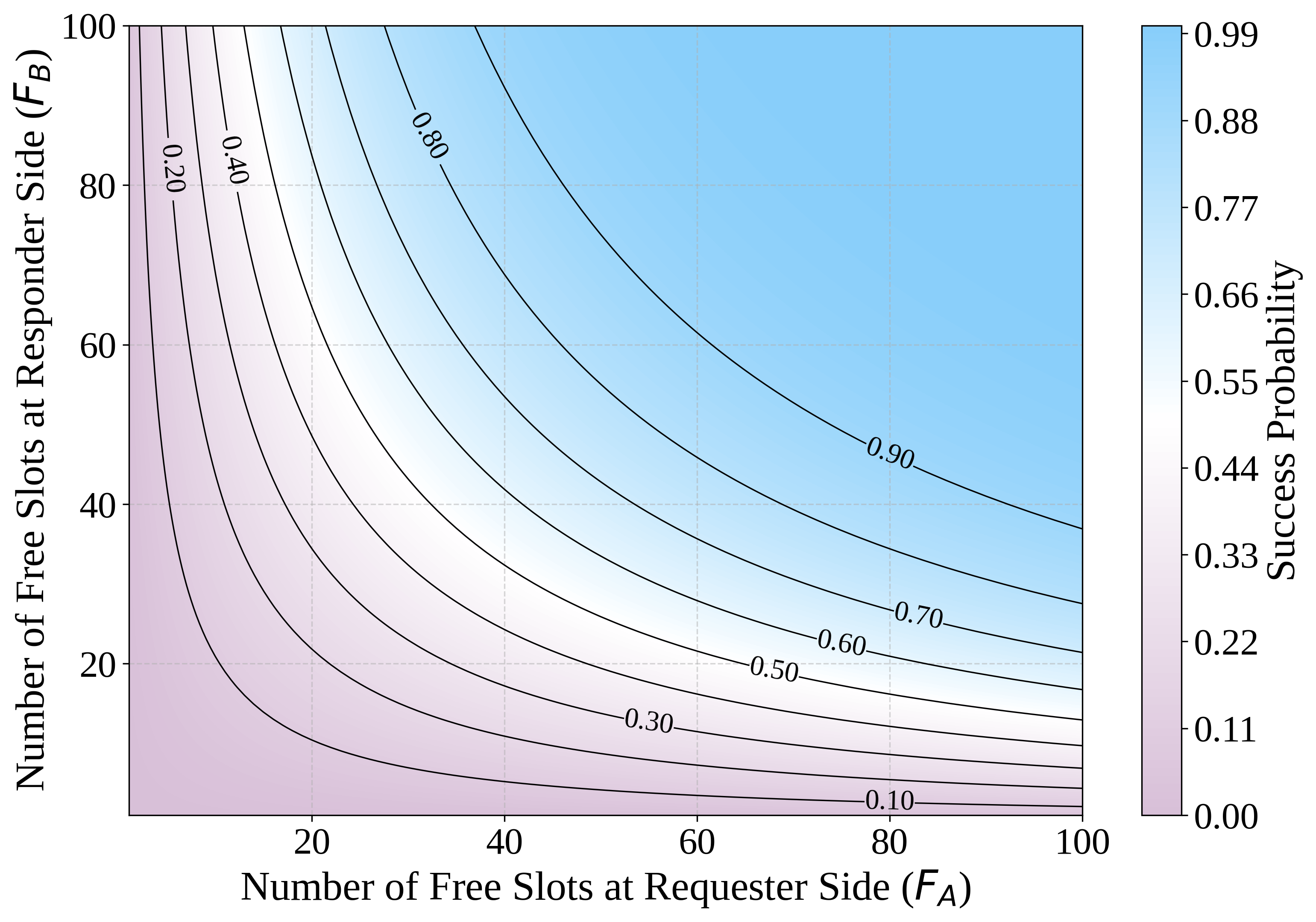}
            \caption{$k=5$}
            \label{fig:success_proba_5}
        \end{subfigure}
        \begin{subfigure}{0.5\linewidth}
            \includegraphics[width=\linewidth]{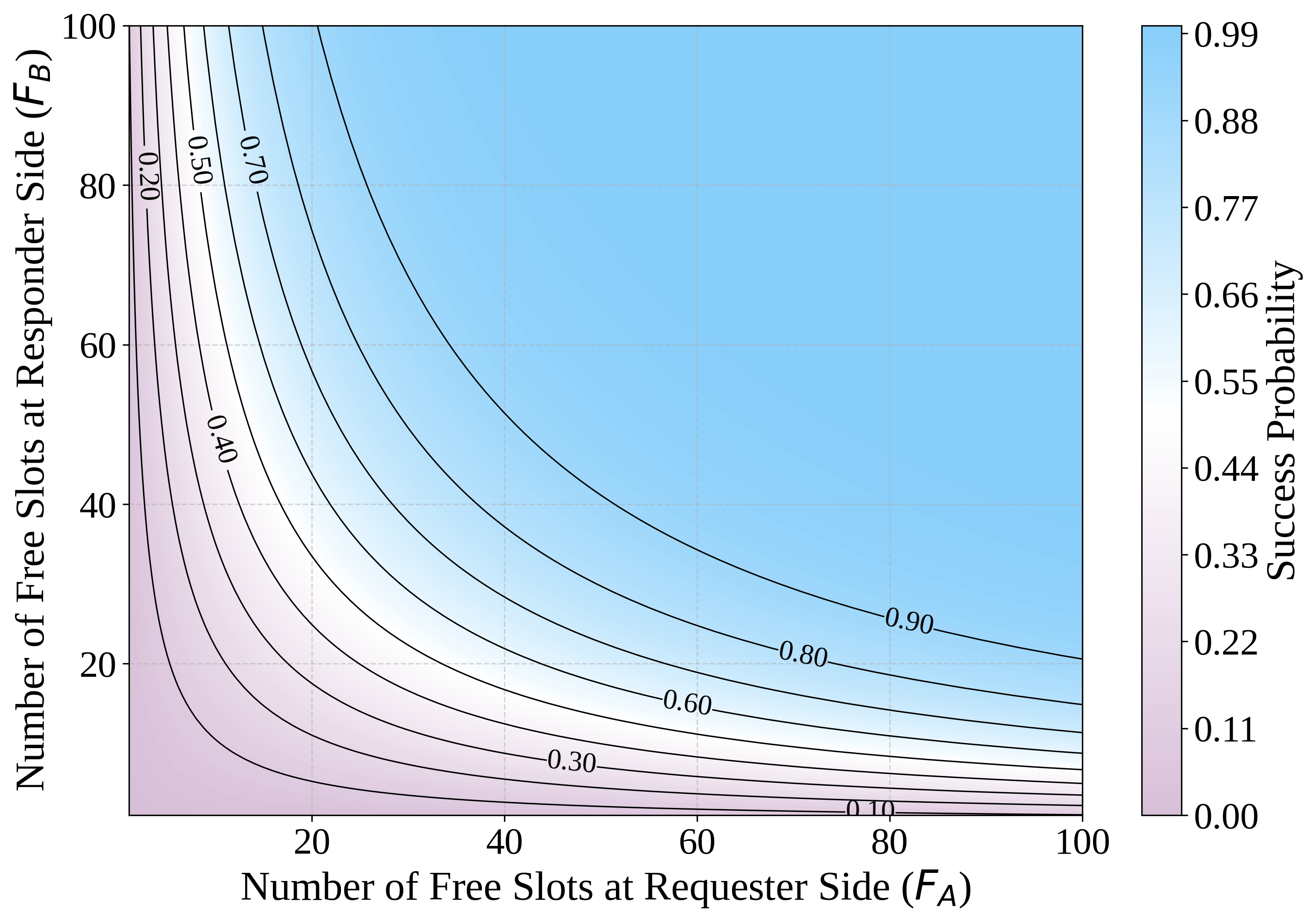}
            \caption{$k=10$}
            \label{fig:success_proba_10}
        \end{subfigure}
        \caption{Impact of free slot availability on first-attempt reservation success probability.}
        \label{fig:success_proba}
    \end{figure}
            
            \item \textit{Inefficient Management of Time Slot Scheduling:}
            {}{In 6TiSCH networks, when a node initiates a cell reservation process using the 6P protocol, one of the two nodes (either the requester or responder, depending on the transaction type) proposes $k$ candidate cells for potential allocation. To prevent scheduling conflicts, the proposing node temporarily locks these cells until it receives a response. As illustrated in Fig. \ref{fig:6p}, this cell-locking phase ensures negotiation consistency but may lead to the underutilization of time slots during the reservation process. The locking period can be prolonged in situations involving packet retransmissions or communication delays, further impacting the slotframe efficiency. Moreover, even when the joining node’s slotframe is fully available, the parent node’s slotframe may be heavily occupied due to ongoing communications with other nodes. This issue intensifies during a parent switching where both nodes may have pre-reserved slots, thus restricting the chance of finding common free cells. As slotframe utilization increases, the probability of successfully allocating at least one common free cell on the first attempt decreases, causing multiple reservation attempts and increased setup delays. 
            The probability of successfully reserving at least one common free cell in the first attempt can be given by:
            \begin{equation}
            \label{eq:Probsucc}
                P_{success} = 1 - \left( 1 - \frac{F_A}{C}* \frac{F_B}{C} \right)^k,
            \end{equation}
            where $F_A$ and $F_B$ are the numbers of free cells at the proposing and responding nodes, respectively, $C$ is the total number of available time slots in the slotframe, and $k$ is the number of proposed cells. Eq.(\ref{eq:Probsucc}) models the probability of at least one successful overlap among the $k$ proposed cells, considering that each time slot has a probability $\left(\frac{F_A}{C}* \frac{F_B}{C}\right)$ of being free at both nodes. 
            For clarity, we present in Fig. \ref{fig:success_proba} how the success probability increases with the number of free cells on the requester and responder sides. Achieving a success probability above 90\% typically requires both nodes to have over 50\% of their slotframes free. To evaluate the impact of the number of proposed cells, we analyze scenarios with $k=5$ and $k=10$, with the choice of $k=5$ being consistent with typical configurations described in RFC 9033 \cite{rfc9033}. When $k=5$, reaching high $P_{success}$ requires large numbers of free cells on both sides. When $k=10$, achieving a similar $P_{success}$ as for $k=5$ requires a lower number of free cells by 35–40\%. Nevertheless, such an improvement comes at the expense of locking more cells during negotiation, potentially causing slotframe underutilization. 
            }

        \item \textit{Inefficient Energy and Time Consumption during 
        Parent Switching:}
			{}{Parent switching in 6TiSCH networks is a critical process that can significantly impact energy consumption and network stability if not properly coordinated with slot allocation.} During parent switching, the RPL standard process ignores the new parent’s schedule {}{status}, thus {}{potentially} leading to an inefficient use of energy and time. {}{This inefficiency stems from the lack of synchronization between the parent selection process handled by RPL and the slot allocation process managed by SF. RPL selects a new parent based solely on routing metrics without verifying if sufficient transmission slots are available. As a result, a node may switch to a parent but fail to send data, causing repeated switching attempts, increased control packet exchanges, and unnecessary energy consumption. Also, queued packets experience increased latency as they wait until the next successful parent switching attempt. This may be due to parent slotframe saturation, simultaneous parent switching by multiple nodes, and delayed slot allocation during cell negotiation. However,} in IIoT systems, where devices operate on limited power, the latter's efficient use is crucial to ensure the system's longevity and reliability. 
				
			\item
			{}{\textit{TSCH Queue Packet Drops due to Cell Shortages:}}  
			{}{
            Poor 6TiSCH queue management can significantly degrade network performance, particularly if traffic dynamics are overlooked. Specifically, the TSCH queue can quickly reach capacity, requiring additional transmission cells to accommodate enqueued packets. A mismatch between queue occupancy and the allocated number of cells can lead to packet drops if additional resources are not proactively secured.}    
    \end{enumerate}
    
	In this work, we attempt to address these challenges through the integration of targeted modifications to the standard 6TiSCH protocol, as described in the next section.

    
    \section{Proposed Solution}
	
	Our work extensively focuses on the intricate interplay between RPL and SF and between SF and packet queue management. The core philosophy of our solution is as follows: 1) Reduce the transmission frequency of control packets upon parent switching; 2) Minimize the number of failed parent switching events that result from cell unavailability; 3) Prevent packet dropping events due to queue saturation; and 4) Improve the jitter and latency performances by slot selection. Our approach targets saving valuable energy resources and mitigating unnecessary medium contention, thus enhancing spatial reuse. 
	
	For understanding purposes of our solution, we highlight the following technical terms usage:
	
	\begin{itemize}
		\item Reference to ``DAO'' is explicitly linked to the DAO packet triggered upon parent switching, unless otherwise specified. This specificity arises from our focus on mechanisms and enhancements relevant to the parent-switching events, while the processes involving periodic DAO packets remain unaltered through our approach.
		
		\item Reference to ``DIO'' may refer to either periodic or direct response to a DIS packet, dependent on the context of the mentioned processes or mechanisms.
		
		
		\item ``DAO-ACK'' and ``ACK'' are two distinct message. DAO-ACK denotes a packet transmitted by the DAG root to acknowledge DAO messages for reliable route registration/de-registration. In contrast, a packet ``ACK'', known also as a link-layer or TSCH acknowledgment (shown in Fig. \ref{fig:TSCH-ACK}), is issued upon packet reception \cite{rfc7554}. 		
		
		\begin{figure}
			\centering
			\includegraphics[width=0.7\columnwidth]{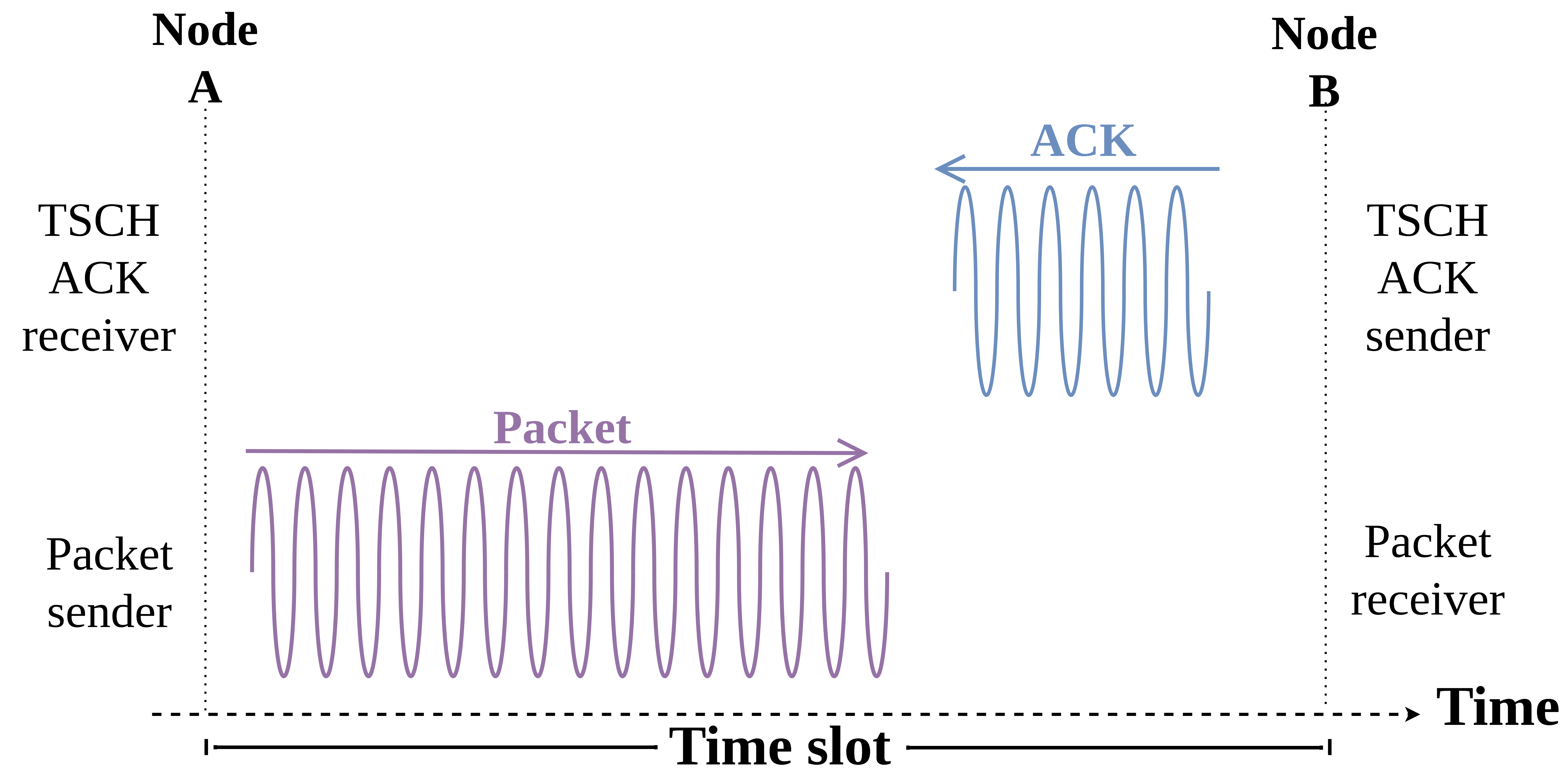}
			\caption{Packet TSCH acknowledgment.}
			\label{fig:TSCH-ACK}
		\end{figure}
		
	\end{itemize}
	
	\subsection{Regulated Triggering of Parent Switching}
	
In 6TiSCH networks, parent switching might be challenging. Indeed, the decision to switch to a potentially superior parent is complex, given the node's lack of awareness regarding the new parent's cell availability. This uncertainty can lead to inefficient energy usage since the node may continue to control packet transmissions without realizing effective cell allocation. Also, the RPL specifications \cite{rfc6550,rfc6552} did not detail the protocol's behaviour regarding transmission interruption to the former parent during this process. Consequently, increased latency and potential packet loss may occur.
	To tackle this issue, our solution proposes to modify the DIO packet to carry a list of available slots of the sender. This approach ensures that a node initiates parent switching only when a shared available slot exists, hence improving the efficiency of the process. Algo. \ref{algo:dio_receiver} below delineates this mechanism, particularly focusing on the node's decision-making process at the juncture of parent switching.
	
	\begin{algorithm}
		\DontPrintSemicolon 
		\KwData{
			$parents$: DIO's senders \\
			$min\_nb$: Minimum number of slots to allocate \\
			$max\_nb$: Maximum number of slots to allocate
		}
		\KwResult{
			Parent switching decision \\
			$nb$: Number of slots to reserve
		}
		
		\tcp{------ Initialization ------}
		$nb\_parentSlot \leftarrow \text{Number of slots allocated to parent}$\;
		$child\_slots \leftarrow \text{Node's free slots}$\; 
		
		\tcp{------ Parent Selection and Slot Reservation ------}
		$selected\_parent \gets \text{parent with minimum rank and available free slots}$\;
		
		\eIf{$child\_slots \geq min\_nb$}{
			\eIf{$\text{Old parent is None}$}{
				\tcp{Node is joining the DODAG}
				$nb \leftarrow min\_nb$\;
			}{
				\tcp{Node is switching parent}                    
				$nb \leftarrow nb\_parentSlot$\;
				\If{$nb > max\_nb$}{
					\tcp{Ensure selected slots are not exceeding the max limit}
					$nb \leftarrow max\_nb$\;
				}
			}
			$free\_slots \leftarrow child\_slots \cap parent\_free\_slots$\;
			$slots \gets \text{Select nearest} \ nb \ \text{slots from} \ free\_slots$\;
			$child\_slots \leftarrow child\_slots - slots$\;
			\text{Append $slots$ to DAO packet}\;
			\text{Append $child\_slots$ to DAO packet}\;
			\text{Transmit DAO packet}\;
		}{
			\text{Pass}\;
			\tcp{Retain current situation since there is no slot availability}
		}
		
		\caption{Parent switching procedure}
		\label{algo:dio_receiver}
	\end{algorithm}

	
	\subsection{Minimizing 6P Transaction Packets}
	\label{sec:minimize}
	In the sequenced control packet transmission process of 6TiSCH networks, depicted in Fig. \ref{fig:network_formation}, when a node decides to change its parent, it communicates this transition by transmitting a DAO packet to the DAG root and triggers the cell reservation process with the new parent.
	This entails dispatching a cell reservation request using a 6P transaction as shown in Fig.\ref{fig:6p}. 
	
	Rather than dedicating the entire packet and the time slot for this purpose, we propose to judiciously embed the requisite information for cell reservation within the RPL control packet DAO. Since the next hop for each child packet is inherently its parent, all packets, including the modified DAO, will go through the designated parent and hence can extract the embedded information about cell reservation and define its relevant cells. This approach is detailed as follows:
	\begin{enumerate}	
		\item \textit{Autonomous channel selection:} Transmission channel {}{$cell_{\rm channel}$} is autonomously determined based on the MAC address of the receiver {}{using a transformation function $f$}, thereby enabling nodes to distribute only the slot offset to represent the cell. {}{$cell_{\rm channel}$ is defined during the reservation using the equation below, which follows the same method used to determine the channel of an autonomous cell:
		\begin{equation}
			cell_{\rm channel} = f(MAC_{\rm address}) \ \%  \ nb\_{\rm channels},
			\label{eq:cell_channel}
		\end{equation}}
		{}{where $nb_{\rm channels}$ is the total number of channels,  and $\%$ is the remainder of the Euclidean division.}
		To avoid interference and benefit from multi-channeling, we adopt the same approach as MSF. Specifically, the current transmission/reception channel ($channel\_{\rm ID}$) is determined based on the Absolute Slot Number (ASN) and the channel offset ($cell_{\rm channel}$) of the current cell as follows:
		\begin{equation}
			channel\_{\rm ID} = \left(ASN + cell_{\rm channel} \right) \ \%  \ nb\_{\rm channels}.
			\label{eq:channel}
		\end{equation}
		
		\item \textit{RPL packet size management:} In 6TiSCH networks where the packet size is intrinsically constrained, judicious management of the RPL packets DIO, DAO, and DAO ACK, is important. Consequently, we propose the following modifications: 
		
		\begin{enumerate}
			\item \underline{\textit{Selective data injection:}} Conventionally, cells are represented as tuples of slot-offset and channel-offset. Their injection into the RPL packets leads to increased packet size and processing time. 
			Aiming to reduce packets' size, we propose a simplified integer-only representation of the available cells instead of using space-greedy tuples.

			\item \underline{\textit{Slot ID encoding:}}
			We propose to implement an 8-bit encoding scheme for slot IDs, thus minimizing the size of the slot ID list injected into the packet.

			\item \underline{\textit{Compact free slots notation:}} Within our proposed approach, the DAO and DIO packets have to convey the node's free slots. To optimize memory usage, we propose that the node transmits either an ordered list of occupied slots or of free slots, depending on which is shorter, and anticipating it with a flag (1 bit). Within the 6TiSCH context, slot ``0'' is reserved by nodes upon joining the network, hence, it will never be included in the ordered list of free slots. This reserved status allows us to efficiently utilize slot ``0'' as the flag for the occupied/free slots list. 
            {}{Since each joined node reserves at least three cells (for Minimal cell, Autonomous Rx, and Autonomous Tx), The largest size of the list that will be injected is
            \begin{equation}
                L_{Length} = \frac{Slotframe_{Length}}{2} -3,
                \label{eq:bigest_list_length}
            \end{equation}}
        	{}{where $Slotframe_{Length}$ is the size of the slotframe (in number of time slots).}
			This flexible method guarantees less memory usage.
		\end{enumerate}
		
		The aforementioned techniques align with the fundamental requirements of LLNs since they aim to reduce transmission overhead, save energy, and enhance the network's throughput. Moreover, they simplify demand processing, which is critical for devices operating within constrained environments, such as in 6TiSCH networks. 

		\item \textit{TSCH ACK-based optimization of slot allocation:}
		For efficient time slot allocation, we propose a novel method that uses TSCH ACK for immediate confirmation of slot reservations upon DAO packet reception. Doing so bypasses the conventionally incurred delays for reservation confirmation. Leveraging TSCH ACK ensures quicker communication of available time slots, thus reducing the locking duration of time slots (shown as $T_{\rm Lock}$ in Fig. \ref{fig:6p}) and increasing the network's flexibility for topological changes. Our technique embeds a list of 
        {}{selected} slots (identified with integers) directly into the TSCH ACK messages. 
        We limit the maximum number of time slots allocated at once to $5$ resulting in an additional payload of merely $5$ bytes in TSCH ACK. Moreover, to 
        {}{enable fast and timely slot selection within the strict time constraints of TSCH}, we assume that each node maintains an up-to-date variable containing its ordered available slots list. This variable is dynamically updated following each slot allocation or de-allocation. Upon receiving {}{the DAO }packet, the node compares the list of available slots embedded within the packet against its variable of available slots. Hence, the decision to reserve slots is based on the most current network status, which improves the responsiveness and adaptability of the 6TiSCH network to real-time demands.

	\end{enumerate}
	
	\subsection{Reducing DODAG Joining Time}
	To reduce packet collisions {}{during network formation} and the nodes' joining time, we propose that the DIO sender sets temporary time slots, ensuring it is actively listening, and it designates one permanent cell to maintain its ongoing listening state. {}{In this way, nodes that receive the DIO packets and attempt to join the DODAG, transmit the DAO packets during the proposed slots, with the channel defined according to Eq.(\ref{eq:channel}). When multiple nodes attempt to send the DAO packet to the DIO sender, they compete for multiple available cells instead of a single cell, thus reducing contention and improving transmission efficiency. 
	To streamline network formation, the proposed mechanism dynamically adjusts the allocation of temporary slots. Initially, a high number of temporary slots, defined as ``\textit{Number of proposed slots per DIO}'', are made available for allocation to accelerate node integration and reduce joining time. These slots are updated with each DIO transmission, ensuring efficient resource management. The DIO sender remains in listening mode during these slots for a duration of ``\textit{DIO cells duration}'', after which it listens only to the permanent slot. The receiver of the DIO packet selects a portion of these slots to send the DAO packet, ensuring controlled channel access and minimizing contention.}

	\subsection{Scheduling Packet Update}
        \label{sec:shedule_enqueue}
\begin{figure}
		\centering
		\includegraphics[width=0.7\columnwidth]{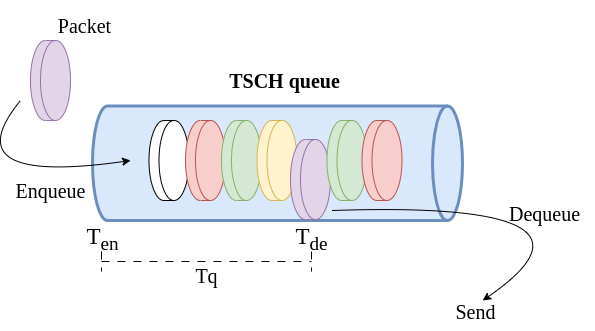}
		\caption{Illustration of a packet time in a 6TiSCH queue.}
		\label{fig:tsch_queue}
	\end{figure}

	In the context of TSCH, transmitting packets from higher-layer protocols is not an instantaneous process; rather, it entails a meticulous queuing and scheduling mechanism. As depicted in Fig. \ref{fig:tsch_queue}, incoming packets are first enqueued within the TSCH queue and await their transmissions. 
	Since we inject the 
    {}{list of} time slots in DIO and DAO packets, they will be locked during time $T_q=T_{\rm de}-T_{\rm en}$, where $T_{\rm en}$ is the instant at which the packet entered the TSCH queue, and $T_{\rm de}$ is the time at which it left the queue. To avoid locking the 
    {}{list of} time slots for a long time $T_q$, we propose to inject them within the packet only before leaving the queue, i.e., at instant $T_{de}$, hence $T_q \approx 0$. 
	{By carefully orchestrating the RPL control packets' update, through adding details about the schedule state (i.e., the free and reserved slot IDs), we optimize time slot utilization.}
	
	\subsection{Avoiding TSCH Queue Overflow}
	
	In 6TiSCH networks, effective synchronization between SF and TSCH queue management is essential for network operation. With various protocols filling the queue with packets, timely processing during the SF-designated transmission cells is crucial. A mismatch between the queue and cell allocation can lead to packet loss. To address this issue, we propose the introduction of an early cell reservation technique. Specifically, as the TSCH queue begins to be filled and before reaching its limit, our system proactively initiates 6P transactions with the preferred parent for an additional cell reservation. This proactive approach offers several benefits: 1) \textit{Prevent packet drops:} Early reservation of additional cells reduces the likelihood of packet loss due to queue overflow, and 2) \textit{Mitigate resource urgency:} This method allows nodes to secure necessary resources in advance, thus avoiding the last-minute pressure of queue overflow and network instability.

	\subsection{Adaptive Self-Sufficient Cell Reservation}
	
\begin{figure}
	\centering
	\includegraphics[trim={0.2cm 0.1cm 1cm 0},clip,width=0.95\textwidth]{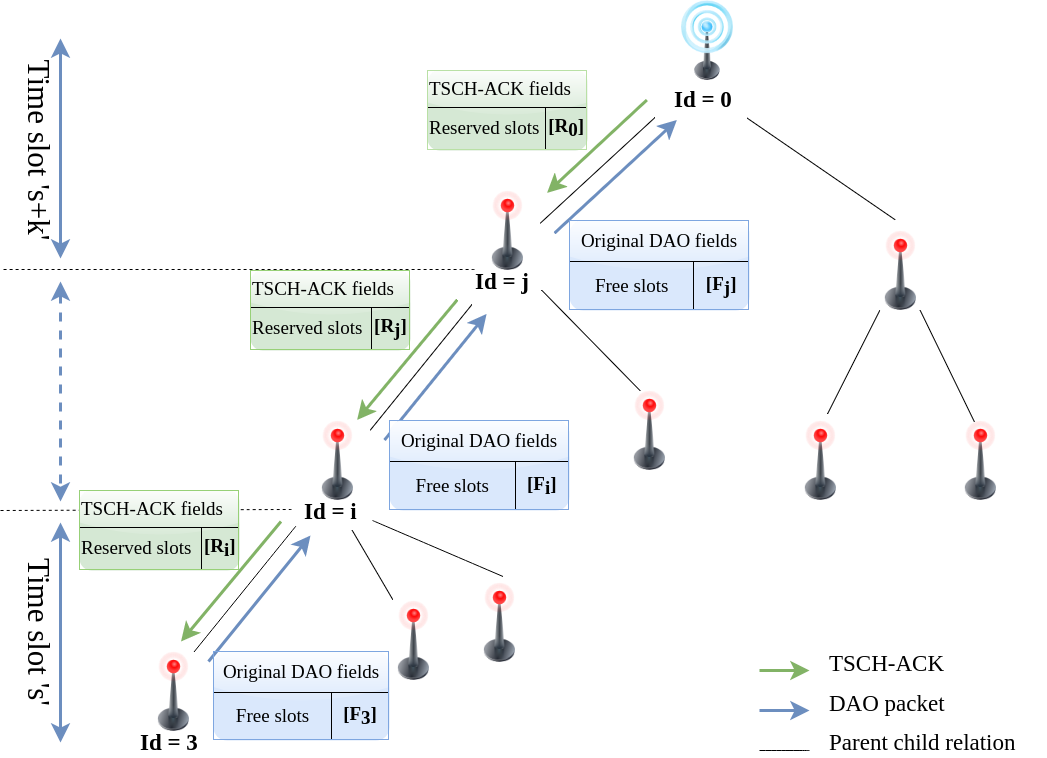}
	\caption{Illustration of the DAO packet transmission and TSCH ACK utilization.}
	\label{fig:DAO-ACK}
\end{figure}

	In a typical 6TiSCH network, all DAO packets are destined towards the DAG root, thus navigating in a multi-hop fashion. In our proposal, essential information about cell reservation, which consists of the available slots and proposed slot lists, is embedded within these DAO packets. When a parent node receives a DAO packet, as it is the principal conduit to the DAG root, it reserves new cells for its new child where the number of cells should not exceed the maximum number of cells to allocate at once. Then, it loads the list of its free slots (instead of the old one that belongs to the child) and the number of slots to be selected (slots to be allocated to its new child) into the DAO packet. This packet is then seamlessly forwarded, with each intermediate node towards the DAG root executing the same procedure. As shown in Fig. \ref{fig:DAO-ACK}, upon reception of a DAO packet from node $3$, node $i$ undertakes cell selection for reservation and embeds the list of chosen slots into the TSCH ACK packet. It then forwards the enhanced packet to its parent.

	To reduce latency at every hop, we propose that the receiver of a DAO packet selects the nearest time slot to the actual one used for the reception. One way to do it is to rely on the concept {}{which is similar to the concept} of the Low-Latency SF (LLSF) protocol \cite{LLSF}, where the slots are selected according to the gap to the left. 
    However, we propose here selecting the nearest slot to the actual one in any direction. Accordingly, if the node is {}{in the process of} joining the DODAG, 
    {}{it selects the nearest free slot to the actual one to start its activity with minimal delay. This mechanism is equally effective for nodes that have already joined, as they continue using the slot allocated to their parent, ensuring better slot continuity, reducing scheduling overhead, and improving end-to-end transmission efficiency}. Consequently, both energy and latency tied to time slot reservations can be reduced. 
    
	
	\section{Performance evaluation}
	\subsection{Simulation Configuration}
	To comprehensively assess the efficiency and resilience of our proposed scheme, we employed the 6TiSCH simulator \cite{https://doi.org/10.1002/ett.3494}.
    {}{This simulator provides a realistic environment for analyzing key network performance metrics, including packet transmission success rates, queue behavior, congestion management, energy consumption, and parent switching efficiency \cite{6TiSCH_Tutorial}. Also, this tool has been extensively used in previous studies to validate 6TiSCH-based scheduling and routing mechanisms \cite{6TiSCH_Tutorial,s24175608,TAPADAR2024103397, s21093074}}.
    For the sake of thorough analysis, generated plots include outlier results, thus reflecting the full range of observed results.

	Our simulations spanned $30$ minutes and $270$ minutes ($4.5$ hours) and were repeated 100 times across two distinct network configurations:
	\begin{itemize}
		\item \textit{Conventional 6TiSCH configuration:} It consists of the use of the conventional MSF as the scheduling function and OF0 as the RPL objective function \cite{ietf-6tisch-msf-10,rfc6552}. Within the results, this configuration is denoted as ``MSF''.
		\item \textit{Proposed solution-based configuration:} This setup leveraged our proposed solution with the modified MSF and OF0 as the scheduling and objective functions, respectively, and labeled as ``PB'' within the results. 
	\end{itemize}

	We conduct our evaluation by randomly placing  $N_1=50$ (resp. $N_2=100$) nodes within a 1 km$^2$ area for each simulation run. The $x$ and $y$ coordinates of these nodes are generated randomly. We employed the Pister-hack connectivity model \cite{https://doi.org/10.1002/ett.3494}, where each node maintains at least three links to neighbors. Throughout the simulations, the observed routing trees ranged from a depth of $2$ hops to $3$ hops, with a median depth of $2$ hops for the network of $N_1=50$ nodes (resp. a median depth between $2$ and $3$ for the network with $N_2=100$ nodes). Moreover, we assume that the frequency of data transmissions for the application or service is in the values $\{ 5, 15, 45, 60 \}$ seconds.  \\
    
    The remaining simulation parameters are presented in Table \ref{tab:sim_para}.
\begin{table}[t]
	\centering
	\caption{Simulation parameters}
	\begin{tabularx}{0.9\linewidth}{Xc}
		\toprule
		\textbf{Parameters}       & \textbf{Value} \\
		\midrule
		Number of nodes           & 50, 100   \\
		Number of simulation runs & 100 \\
		Simulation duration (min) & 30, 270  \\
		Network area (km$^2$) & 1 \\
		Service or application packet period (sec) & 5, 15, 45, 60  \\
		Slotframe length (in number of slots) & 100  \\
		Time slot duration (sec) & 0.01 \\
		Secure Joining required & True \\
		TSCH queue size (number of packets) & 10 \\
		Number of radio channels & 16 \\
		 & \\
		\begin{tabular}{@{}l@{}} 
			\\
			TSCH maximum payload length (bytes) \\
			DIO packet size (bytes) \\ 
			DAO packet size (bytes) 
		\end{tabular} &
		\begin{tabular}{@{}c|c@{}} 
			MSF & PB \\ \midrule
			90  & 120 \\
			76  & 120 \\
			20  & 75
		\end{tabular} \\
		\bottomrule
	\end{tabularx}
	\label{tab:sim_para}
\end{table}

    
    {}{\subsection{Parameters' Values Setup}}
    {}{\subsubsection{DIO and DAO Packet Size Selection}}
    {}{Based on eq.(\ref{eq:bigest_list_length}), with a slotframe size $Slotframe_{Length}=100$ slots, the largest size of the free slots' list is 47 slots. Since each slot ID is encoded on 1 byte, this requires adding 47 bytes to a packet. To accommodate this additional information while ensuring the packets remain within the TSCH maximum payload length (i.e., 120 bytes), we extend the conventional DIO and DAO packet sizes from 76 and 20 bytes to 120\footnote{{}{With the additional 44 bytes for the DIO, a maximum of 44 slot IDs can be injected.}} and 75 bytes, respectively.
    } 
    
    


    {}{\subsubsection{Queue Management Parameters}}
        
	{}{Upon each packet enqueuing operation, nodes assess the remaining capacity within the TSCH queue. When the available space in the queue is less than or equal to ``\textit{TSCH queue empty places threshold}'', the node initiates a cell reservation request using the 6P protocol. It will request ``\textit{Number of cells reserved upon overflow}'' cells, and subsequent request is constrained by the value of the parameter ``\textit{Cycle interval before next cell request}''.}
	
	{}{A low value of ``\textit{TSCH queue empty places threshold}'' may prevent the mechanism from functioning effectively. Conversely, a higher threshold may lead to extensive cell reservation requests, thus increasing the number of allocated cells beyond what is necessary. While this may accelerate packet transmission, it can also result in inefficient resource utilization. The delay of ``\textit{Cycle interval before next cell request}'' is essential to avoid multiple consecutive reservation requests. However, its value must be carefully balanced. A low value may not allow enough time to resolve the previous overflow before triggering another request, leading to multiple cell reservation requests, which can increase scheduling overhead and inefficient resource allocation. In contrast, a long period may cause the node to overlook queue overflows, delaying necessary cell reservations and potentially increasing packet loss.}

    \begin{figure}
        \begin{subfigure}{0.5\linewidth}
            \includegraphics[width=0.98\linewidth]{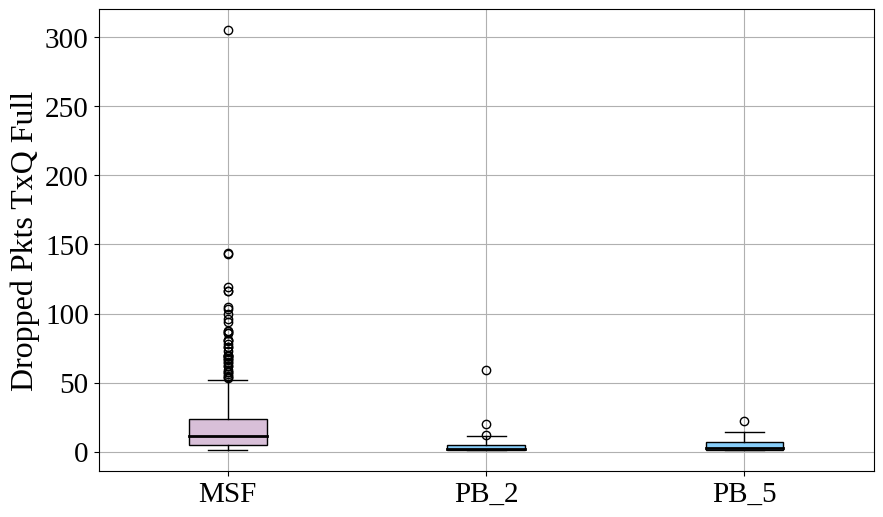}
            \caption{$N=50$}
            \label{fig:Queue_full_drop_50_nodes}
        \end{subfigure}
        \begin{subfigure}{0.5\linewidth}
            \includegraphics[width=\linewidth]{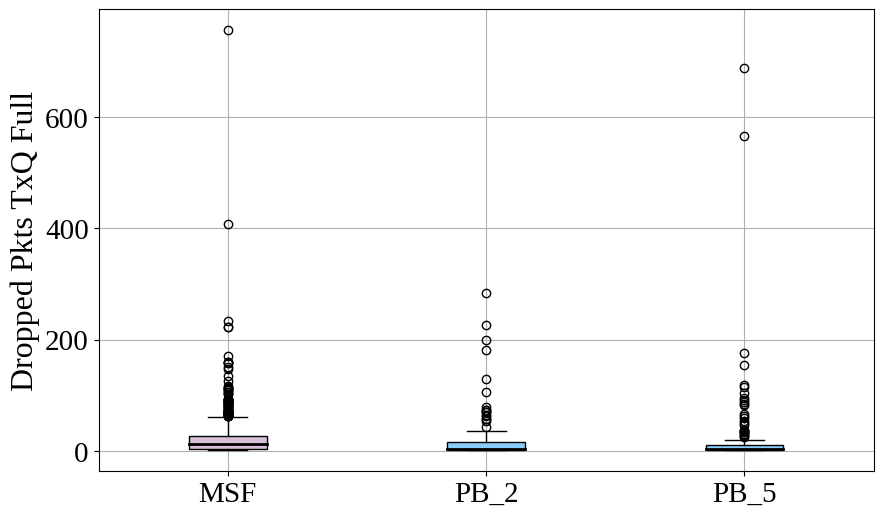}
            \caption{$N = 100$}
            \label{fig:Queue_full_drop_100_nodes}
        \end{subfigure}
        \caption{Number of dropped packets due to queue overflow vs. size of ``Cycle interval before the next cell request''.}
        \label{fig:nb_queue_dropped}
    \end{figure}

    {}{To define the value of parameter ``\textit{Cycle interval before next cell request}'' and demonstrate the performance of our mechanism, we conducted simulations with an application period of 5 seconds. This setup models high-traffic conditions, where queue congestion is most significant and the effectiveness of queue management mechanisms becomes more evident. In Fig. \ref{fig:nb_queue_dropped}, the performance of the MSF benchmark is compared with our proposed solution under two configurations, namely $PB\_2$ and $PB\_5$, representing cycle intervals of 2 and 5 slotframes, respectively. As shown, MSF exhibits a significantly higher number of packet drops due to queue overflow, with notable variability. In contrast, our method, using $PB\_2$ or $PB\_5$, substantially reduces packet losses. With $PB\_2$, our technique demonstrates slightly better performances than for $PB\_5$, highlighting that a cycle interval of 2 slotframes is sufficient to mitigate queue overflow. Increasing the interval beyond 2 slotframes would delay necessary slot allocations, potentially impacting data transmission efficiency. Hence, in the remaining, we select \textit{Cycle interval before next cell request}'' to be equal to 2 slotframes.
    }
  
%
		
	\begin{figure}
		\begin{subfigure}{0.5\linewidth}
			\includegraphics[width=0.94\textwidth]{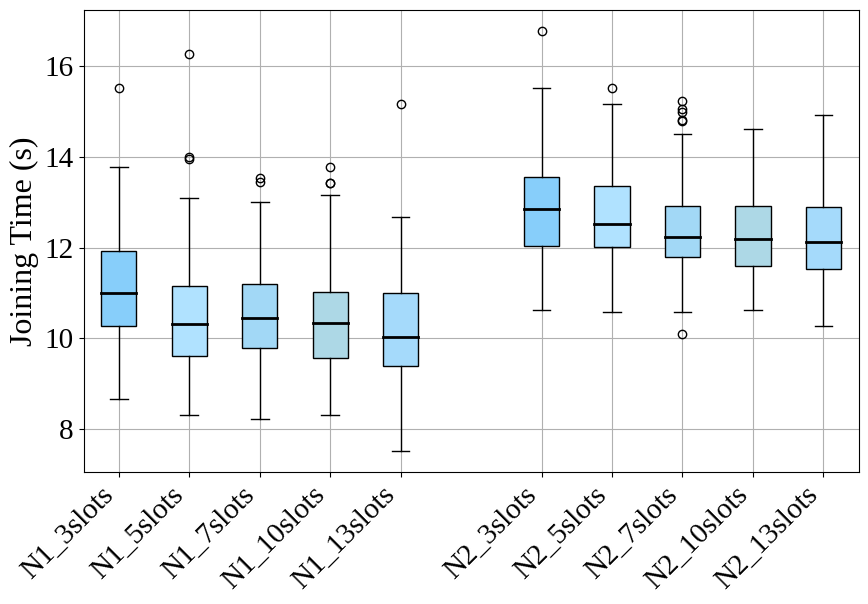}
			\caption{Joining time (in sec)}
			\label{fig:DIO_slots_join}
		\end{subfigure}	
		\begin{subfigure}{0.5\linewidth}
			\includegraphics[width=\textwidth]{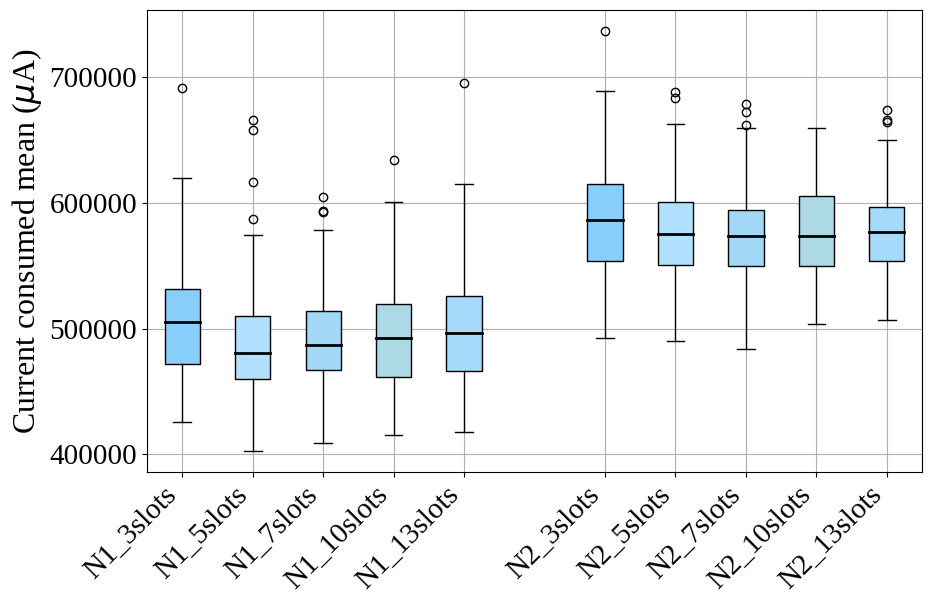}
			\caption{Consumed current (in $\mu$A)}
			\label{fig:DIO_slots_energy}
		\end{subfigure}
		\caption{Joining time and consumed current vs. nbr. of proposed slots per DIO.}
	\end{figure}

    
        {}{\subsubsection{Number of Proposed Slots per DIO}}  
		To set the parameter ``\textit{Number of proposed slots per DIO}'' to the best value, we conducted several experiments with different values. Specifically, in Figs. \ref{fig:DIO_slots_join}-\ref{fig:DIO_slots_energy}, we evaluated the impact of different numbers of proposed time slots per DIO on the joining time and current consumption (equivalent to energy consumption). 

		These experiments have been conducted over an observation time of $30$ minutes, with an application period of $5$ seconds, and for different network sizes, i.e., $N_1=50$ and $N_2=100$ nodes. As shown in Fig. \ref{fig:DIO_slots_join}, for any number of nodes, the joining time reduces when the number of proposed slots is higher. This is expected since more proposed slots mean that a novel node can rapidly be configured and join the DODAG. However, since the number of proposed slots cannot be very high, as it may increase the size of packets, our preference is for the value $7$ since it provides similar results to the system with the value $13$, in particular for $N_2$, i.e., for a large network. A similar explanation can be provided for Fig. \ref{fig:DIO_slots_energy} where the value $7$ achieves low current consumption for the large network. Indeed, a very high value, such as $13$, can result in inefficient idle listening, thus increasing power consumption. Consequently, we recommend the ``\textit{number of proposed slots per DIO}'' to be $7$ as the best value since it is the lowest one that achieves good performances in terms of joining time and consumed power. 	
		

   \begin{figure}
	\begin{subfigure}{0.5\linewidth}
			\includegraphics[width=0.94\textwidth]{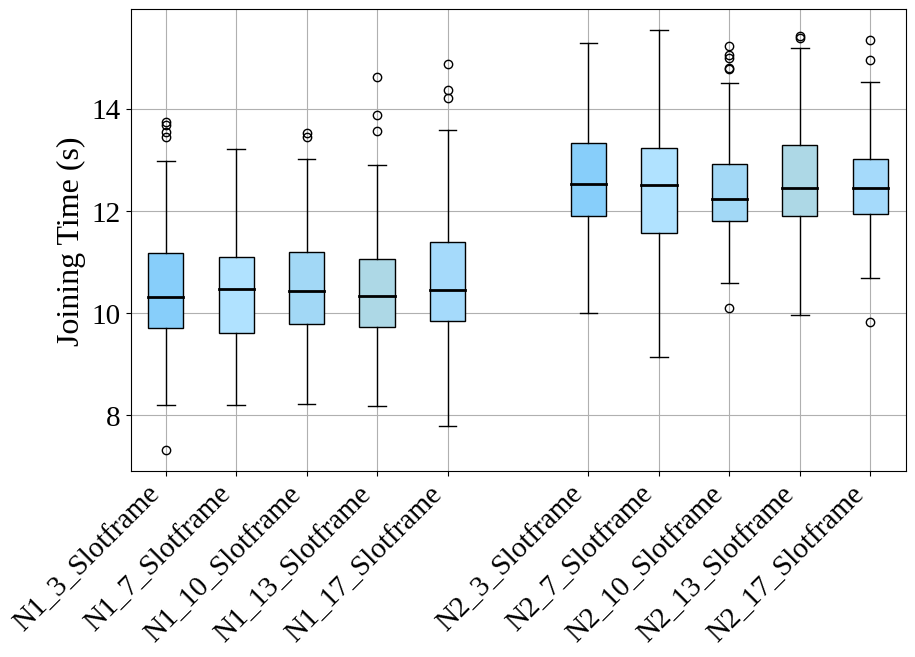}
			\caption{Joining time (in sec)}
			\label{fig:DIO_duration_join}
	\end{subfigure}	
	\begin{subfigure}{0.5\linewidth}
			\includegraphics[width=\textwidth]{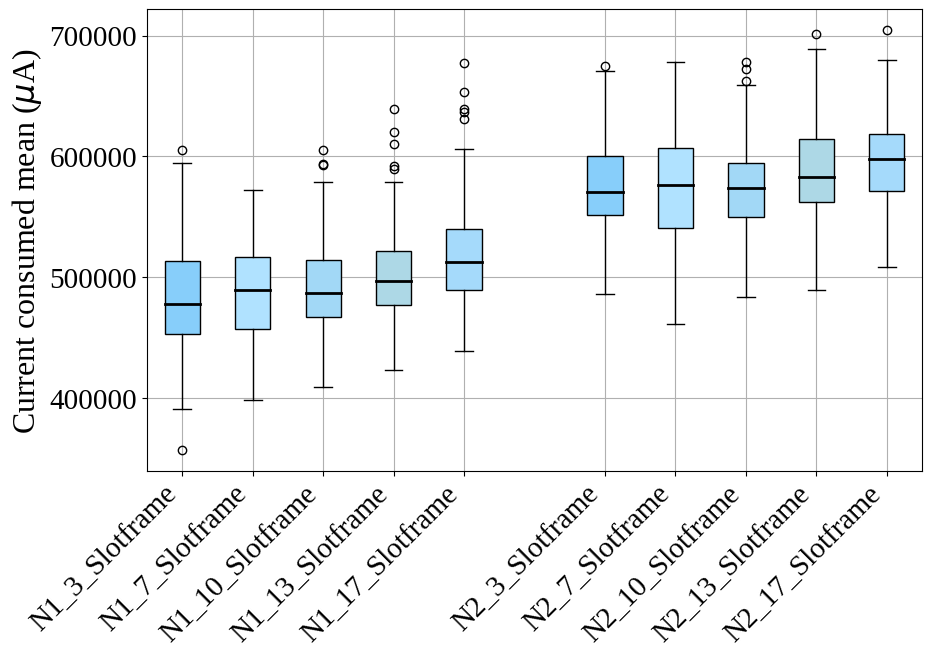}
			\caption{Consumed current (in $\mu$A)}
			\label{fig:DIO_duration_energy}
	\end{subfigure}
		\caption{Joining time and consumed current vs. proposed DIO cells duration.}
    \end{figure}
          
    {}{\subsubsection{DIO Cell Duration Selection}}
    {}{The selection of the ``\textit{DIO cells duration}'' parameter (expressed in number of slotframes), which defines how long a node maintains proposed cells for new child nodes in its DIO, is based on extensive simulation experiments following the same methodology used in Subsection 5.2.3. The DIO cells duration was tested for different values, i.e., 3, 7, 10, 13, and 17 slotframes, while measuring key performance metrics, i.e., join time and current consumption, in different network sizes $N_1=50$ and $N_2=100$ nodes, and illustrated in Figs. \ref{fig:DIO_duration_join}-\ref{fig:DIO_duration_energy}. These metrics were assessed as this mechanism aims to reduce the joining time while minimizing energy drain.}     {}{According to Fig. \ref{fig:DIO_duration_join}, the joining times are similar, e.g., around 10.5 seconds for $N_1$, for any size of  DIO cells duration. However, when looking at Fig. \ref{fig:DIO_duration_energy}, we notice that energy consumption is the smallest for a duration equal to 10 slotframes. This suggests that a DIO cells duration of 10 slotframes provides the best trade-off between reducing joining time and minimizing energy consumption.}


    


{}{We summarize in Table \ref{tab:pb_conf} the selected parameters for our solution.} 

  \begin{table}[t]
    \small
    \caption{Parameters of the proposed solution}
    \renewcommand{\arraystretch}{1.2}
    \begin{tabularx}{\columnwidth}{>{\raggedright\arraybackslash}Xcccc}
        \toprule
        \textbf{Parameter} & \multicolumn{4}{c}{\textbf{Value}} \\
        \midrule
        Number of proposed slots per DIO & \multicolumn{4}{c}{7} \\
        DIO cells duration (in number of slotframes) & \multicolumn{4}{c}{10} \\
        Number of permanent DIO slots & \multicolumn{4}{c}{1} \\
        Slot selection ratio within DIO slots & \multicolumn{4}{c}{3} \\
        Maximum number of cells to reserve at once & \multicolumn{4}{c}{5} \\
        Initial phase duration (min) & \multicolumn{4}{c}{45} \\
        \cmidrule(lr){1-5} 
        \multicolumn{5}{c}{\textit{Queue Management Parameters}} \\
        \cmidrule(lr){1-5}
        TSCH queue empty places threshold & \multicolumn{4}{c}{2} \\
        Cycle interval before next cell request & \multicolumn{4}{c}{2} \\
        Number of cells reserved upon overflow & \multicolumn{4}{c}{1} \\
        \bottomrule
    \end{tabularx}
    \label{tab:pb_conf}
\end{table}

    \begin{figure}
        \begin{subfigure}{0.5\linewidth}
            \includegraphics[width=0.98\linewidth]{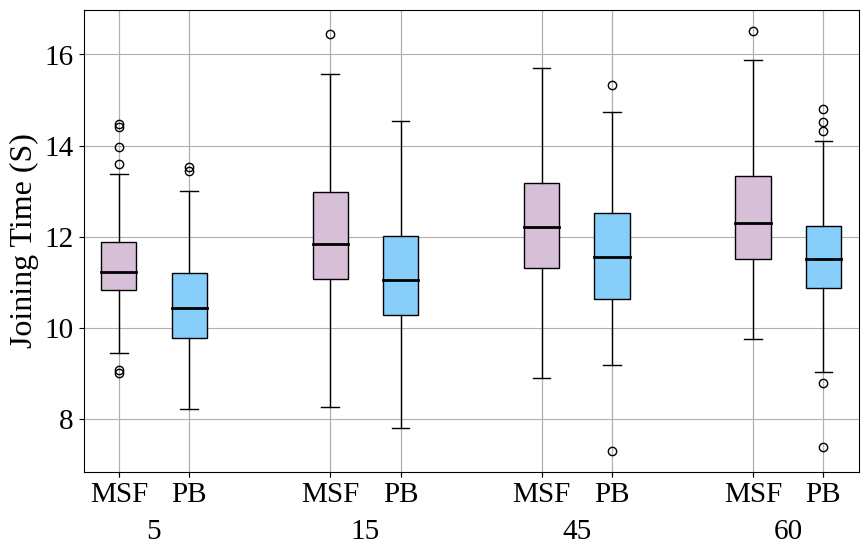}
            \caption{$N=50$}
            \label{fig:join_time_50nodes}
        \end{subfigure}
        \begin{subfigure}{0.5\linewidth}
            \includegraphics[width=\linewidth]{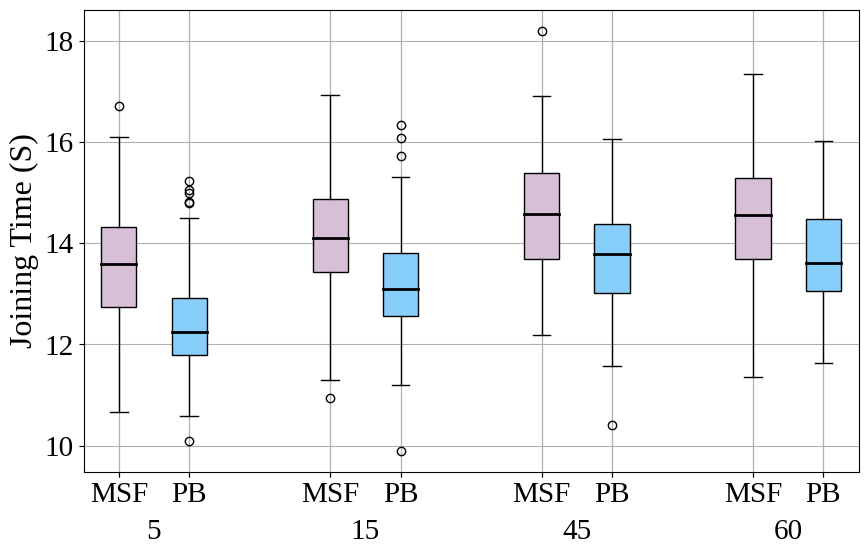}
            \caption{$N = 100$}
            \label{fig:join_time_100nodes}
        \end{subfigure}
        \caption{Joining time (in sec) vs. length of service packet period (short term).}
        \label{fig:join_time}
    \end{figure}
    
    \subsection{Simulation Results}
		
		For our simulations, we define two scenarios, the first is in the short term with a duration of $30$ min, while the second is in the long term with a duration of $4.5$ hours.
		
    \subsubsection{Short Term Results}
		The initial phase of the network's life is known as network formation. This phase is pivotal for establishing the network's parameters and topology. 
		 
		 To demonstrate the effectiveness of our solution in accelerating the network formation timeline and minimizing power consumption, we conducted simulations {}{that last $30$ minutes}.	

		\begin{figure}
			\begin{subfigure}{0.5\linewidth}
				\includegraphics[width=\linewidth]{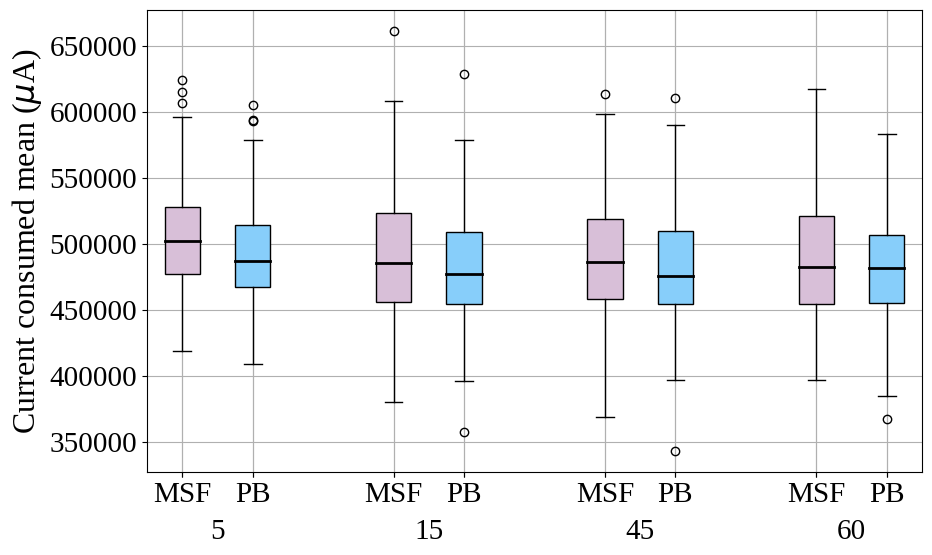}
				\caption{$N = 50$}
				\label{fig:Global_current_30mn_50nodes}
			\end{subfigure}		
			\begin{subfigure}{0.5\linewidth}
				\includegraphics[width=0.95\linewidth]{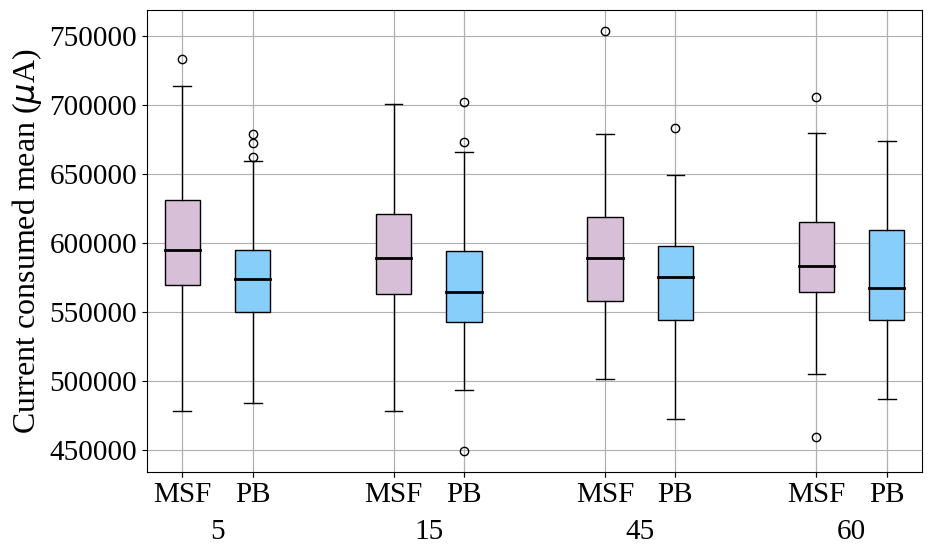}
				\caption{$N = 100$}  
				\label{fig:Global_current_30mn_100nodes}                      	 
			\end{subfigure}
			\caption{Consumed current (in microampere) vs. length of service packet period (in sec) (short term).}
			\label{fig:Global_current_30mn}
		\end{figure}
  
		\begin{figure}
			\begin{subfigure}{0.5\linewidth}
				\includegraphics[width=\textwidth]{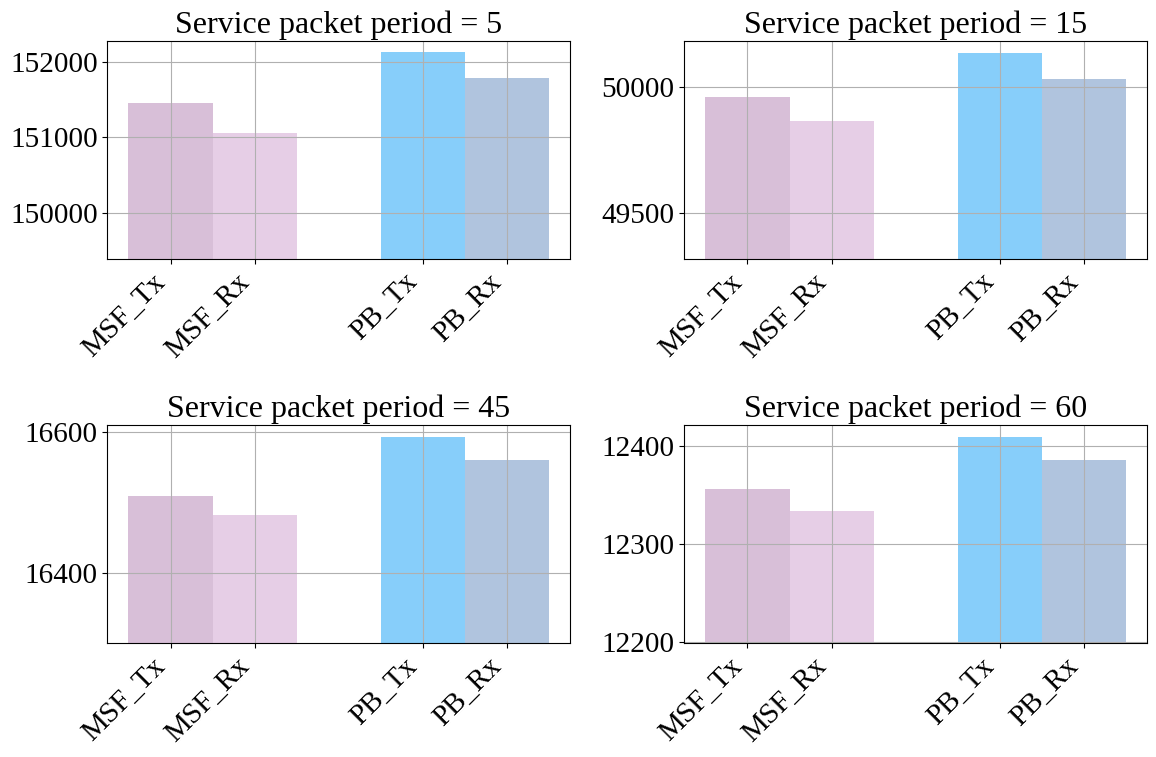}
				\caption{$N = 50$}
			\end{subfigure}	
			\begin{subfigure}{0.5\linewidth}
				\includegraphics[width=\textwidth]{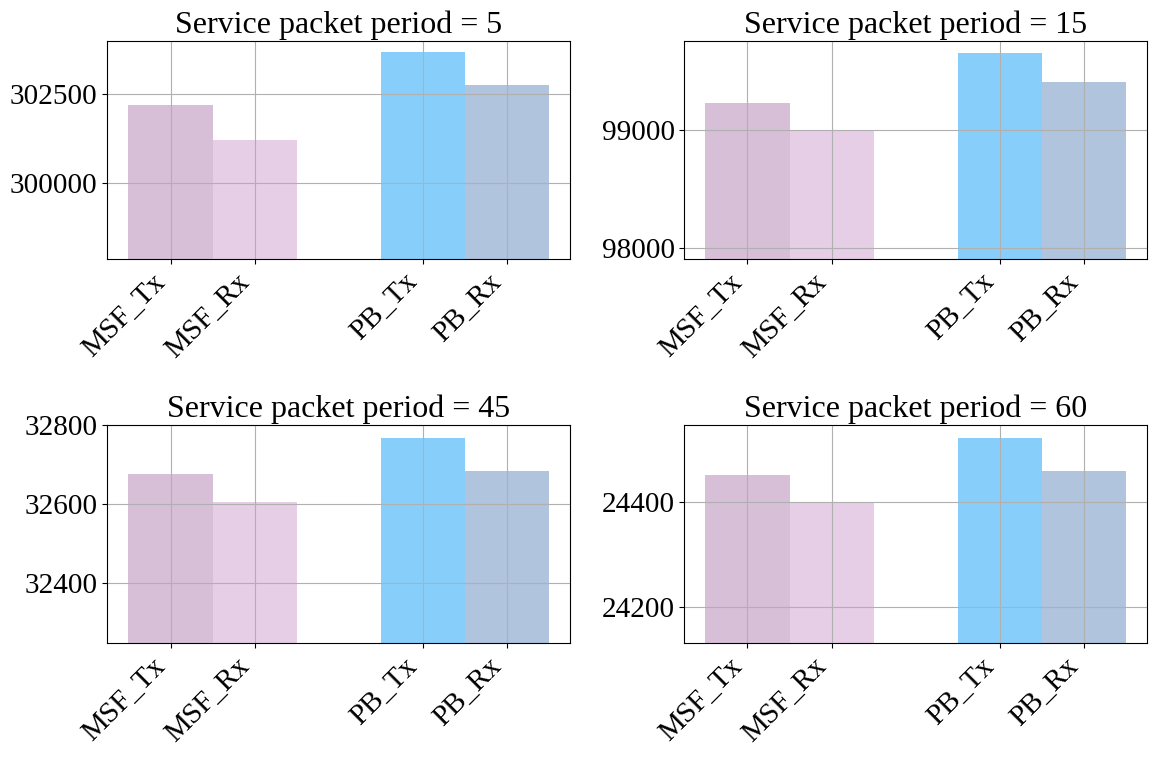}
				\caption{$N = 100$}                        	 
			\end{subfigure}
			\caption{Nbr. of Tx/Rx packets (different service packet periods, long term).}
			\label{fig:total_sent_received}
		\end{figure}
		
		As shown in Fig \ref{fig:join_time}, our proposed solution PB achieves lower joining time for different number of nodes within the network, $N$, and service packet period duration. For instance, PB achieves an average joining time of $11$ seconds for a service packet period equal to $15$ and $N=50$, while MSF needs approximately $12$ seconds in the same conditions.
		Indeed, a node within the PB framework uses proposed slots in the received DIO to define the cells during which it sends the DAO, unlike MSF which relies on sending {}{the DAO packet and the} 6P request to its preferred parent and then waiting for {}{the 6P} response for cell reservation. In PB, the DAO packet serves a dual purpose: Communicating with the RPL to establish the DODAG structure and participating in SF. 
		As the service packet period increases, we notice a degradation in the joining time. This is expected since a node with a low (resp. high) service packet period will transmit EB packets more (resp. less) often, which increases (resp. reduces) the likelihood that other nodes synchronize and join the DODAG.
						
		In 6TiSCH networks, the process of network formation often expends a significant amount of energy, even before useful data transmission. {}{To evaluate energy consumption, we adopt a model based on the operational state of nodes in each time slot, as implemented in the 6TiSCH simulator. Time slots in which the node is desynchronized are not considered as the node is inactive in such cases. The total charge consumption over the simulation period $Q_{\text{total}}$ is computed as the sum of the charge consumed in each state $C_{\text{state}}$ \cite{tsch_energy}, i.e.,	
			\begin{equation}
				Q_{\text{total}} = \sum_{\text{state}} N_{\text{state}} \cdot C_{\text{state}},
				\label{eq:charge}
			\end{equation}			
			where \( N_{\text{state}} \) represents the number of occurrences or total duration spent in each state.} In Fig.  \ref{fig:Global_current_30mn}, we present the consumed current related to the scenarios of Fig. \ref{fig:join_time}. The PB scheme performs better than the MSF one, for any $N$ and service packet period duration. The performance gap increases as $N$ becomes higher. {}{The observed improvement in join time and energy efficiency are due to embedding slot proposals within the DIO packet rather than using 6P transactions. Also, by merging control packets, our method accelerates node activation, enabling newly joined nodes to quickly invite other nodes into the DODAG and enhancing the overall network convergence. As the network scales up, competition for shared resources intensifies leading to increased transmission attempts and current consumption. However, our solution remains particularly robust in such conditions. For instance, in Fig. \ref{fig:Global_current_30mn}, MSF exhibits high current consumption compared to PB in any network size. The ability of PB to proactively allocate resources results in more stable current consumption.} 
		Finally, we notice that the service packet period has a slight impact on the total consumed current. 

        
		\subsubsection{Long term results}
		We assess here the performances of PB and MSF for long-term operation, in terms of the number of transmitted/received packets, end-to-end (E2E) latency, network jitter, consumed charge, and nodes' lifetime. 

	
	Fig. \ref{fig:total_sent_received} illustrates the number of transmitted (Tx) and received (Rx) packets for different service packet periods and $N$. For any $N$ and service packet period, our solution outperforms the MSF benchmark. Indeed, the network that uses our approach realizes more frequent transmissions, and nodes within it successfully receive more packets. When the service packet period increases, i.e., a longer service is running, the number of exchanged packets drops. such a result is expected since during the observation time of 4.5 hours, a longer service time impacts the frequency of its execution over the observation period. Moreover, when $N=100$, the number of exchanged packets almost doubles compared to the case with $N=50$. Indeed, the availability of more nodes joining the DODAG makes it possible to transmit/receive more packets, within the limits of available channels and time slots of 6TiSCH.  

	\begin{figure}
		\begin{subfigure}{0.5\linewidth}
			\includegraphics[width=0.98\textwidth]{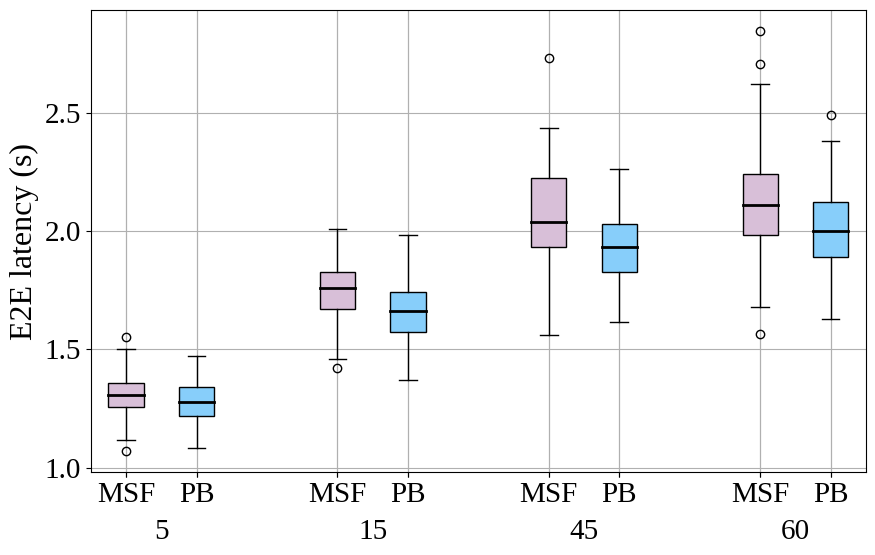}
			\caption{$N = 50$}
			\label{fig:latency_50nodes}
		\end{subfigure}		
		\begin{subfigure}{0.5\linewidth}
			\includegraphics[width=\textwidth]{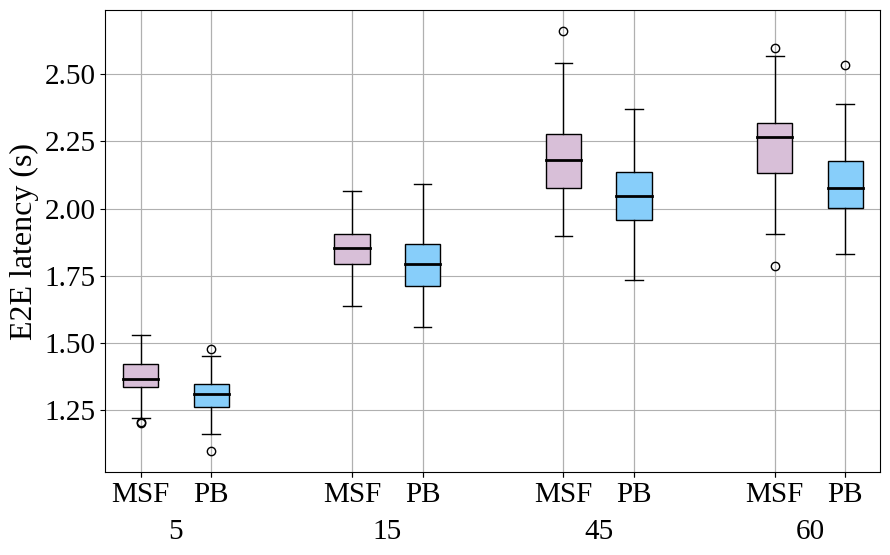}
			\caption{$N = 100$}
			\label{fig:latency_100nodes}
		\end{subfigure}
		\caption{E2E latency (in sec) vs. service packet period (long term).}
		\label{fig:latecny}
	\end{figure}
	
	
	End-to-end latency is defined as the duration measured from the absolute slot number (ASN) at which a source node transmits a service/application packet, to the ASN at which this packet is successfully received and acknowledged by the destination, while jitter refers to the variation in packet transmission latency over time within a network. It is the inconsistency observed in the arrival time of packets, primarily caused by varying network congestion levels, route changes, or other unforeseen factors. Fig. \ref{fig:latecny} shows the network's E2E latency for different service packet periods and $N$. As it can be seen, the PB scheme achieves lower E2E latency in any network condition than the MSF. This result underscores the efficiency of our method. 
	
	\begin{figure}		
	\begin{subfigure}{0.82\linewidth}
		\includegraphics[width=\textwidth]{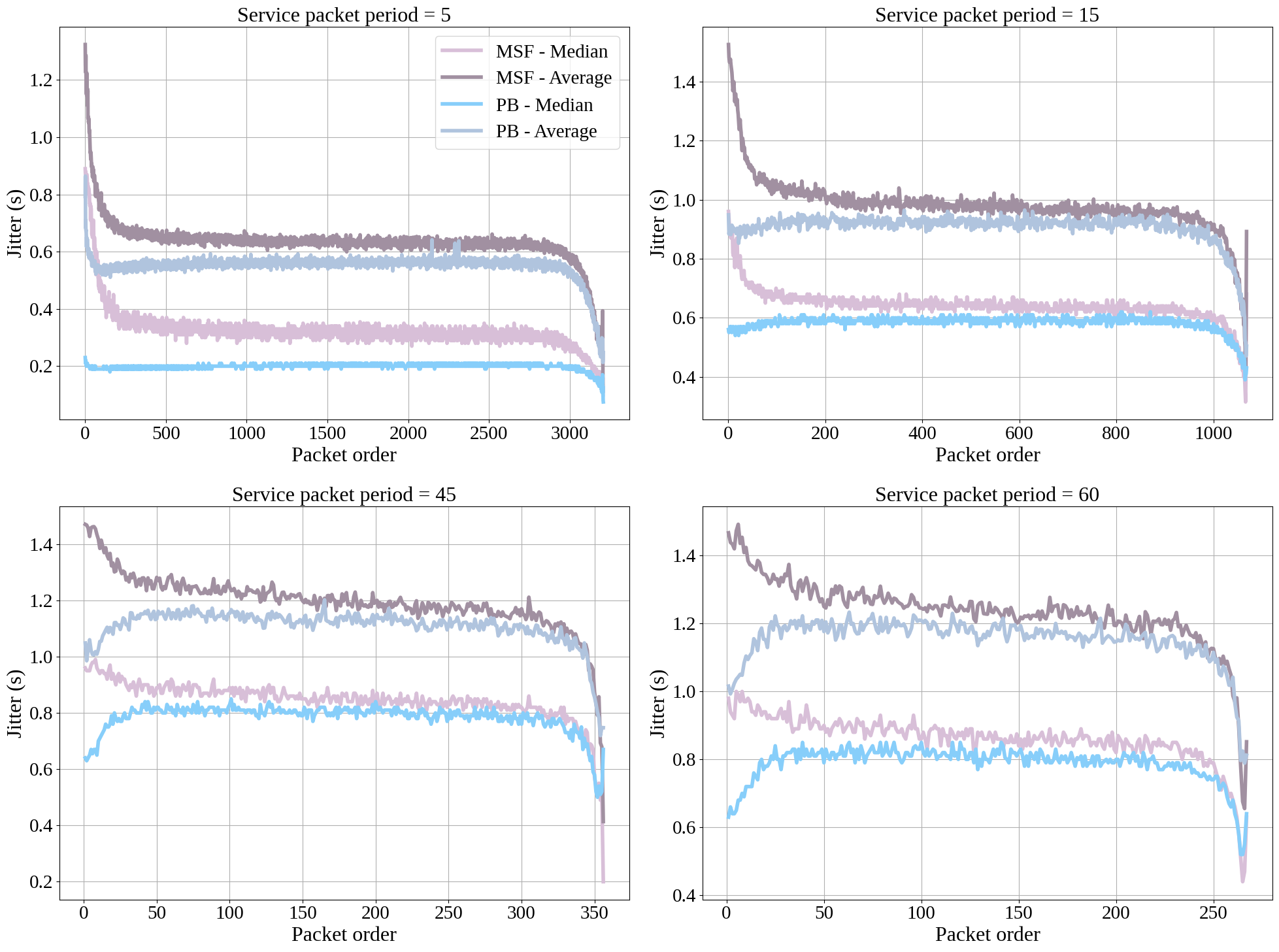}
		\caption{$N=50$}
		\label{fig_N50_jitter}
	\end{subfigure}
	\begin{subfigure}{0.82\linewidth}
		\includegraphics[width=\textwidth]{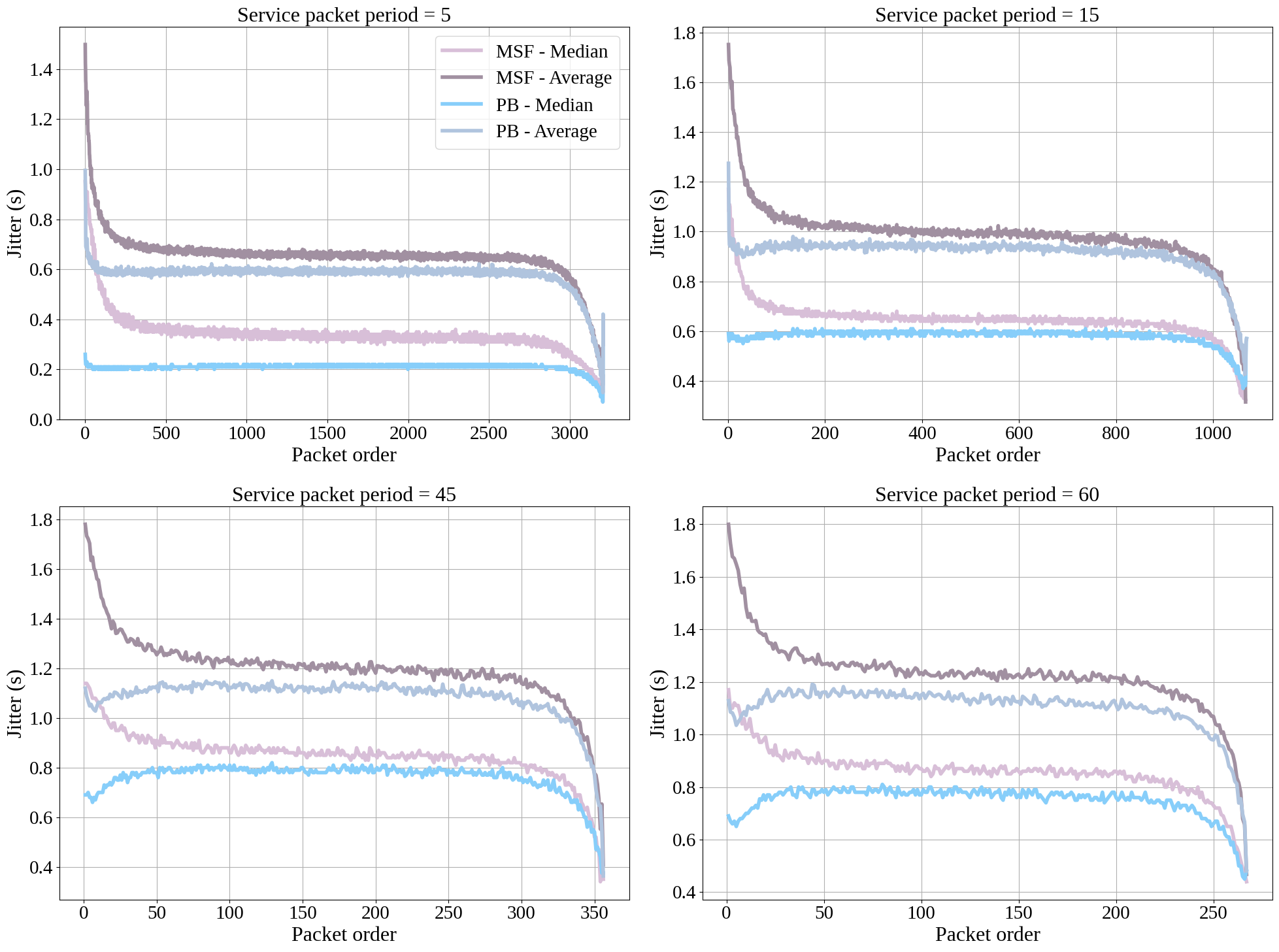}
		\caption{$N=100$}
		\label{fig_N100_jitter}
	\end{subfigure}
	\caption{Median and avg. jitter of Tx packets (different service packet periods, long term).}
	\label{fig_jitter}
	\end{figure}

	In Fig. \ref{fig_jitter}, we plot the median and average jitter performance for the transmitted packets during the observation time, for different $N$ and service packet periods. To obtain the median and average results, we generated $100$ scenarios for each packet order point. The median offers insight into the typical experience of a node without the influence of outliers, while the average jitter aggregates these variations to provide a general sense of the network's jitter over time. According to Fig. \ref{fig_jitter}, MSF shows an initial peak (median and average jitters), due to the network's formation phase, which might disrupt real-time services. In contrast, PB consistently maintains lower jitter performances than MSF, which indicates a stable packet delivery time. The median jitter of MSF exhibits high variability, thus unsuitable for services requiring consistent data rates, while PB follows a more predictable jitter performance, which makes it fit for real-time applications. As $N$ increases from $50$ to $100$, and for a given scheme and service packet period, we notice that the jitter performance is stable and does not degrade. 
    For a given $N$, the jitter degrades with the service packet period. Indeed, a longer service time impacts packet transmissions and thus might cause additional delay and/or instability. In summary, the jitter is slightly impacted by the size of the network, while the influence of the service packet delay is more significant on both the average and median jitters. \\
	
		
	\begin{figure}
		\begin{subfigure}{0.82\linewidth}
			\includegraphics[width=\textwidth]{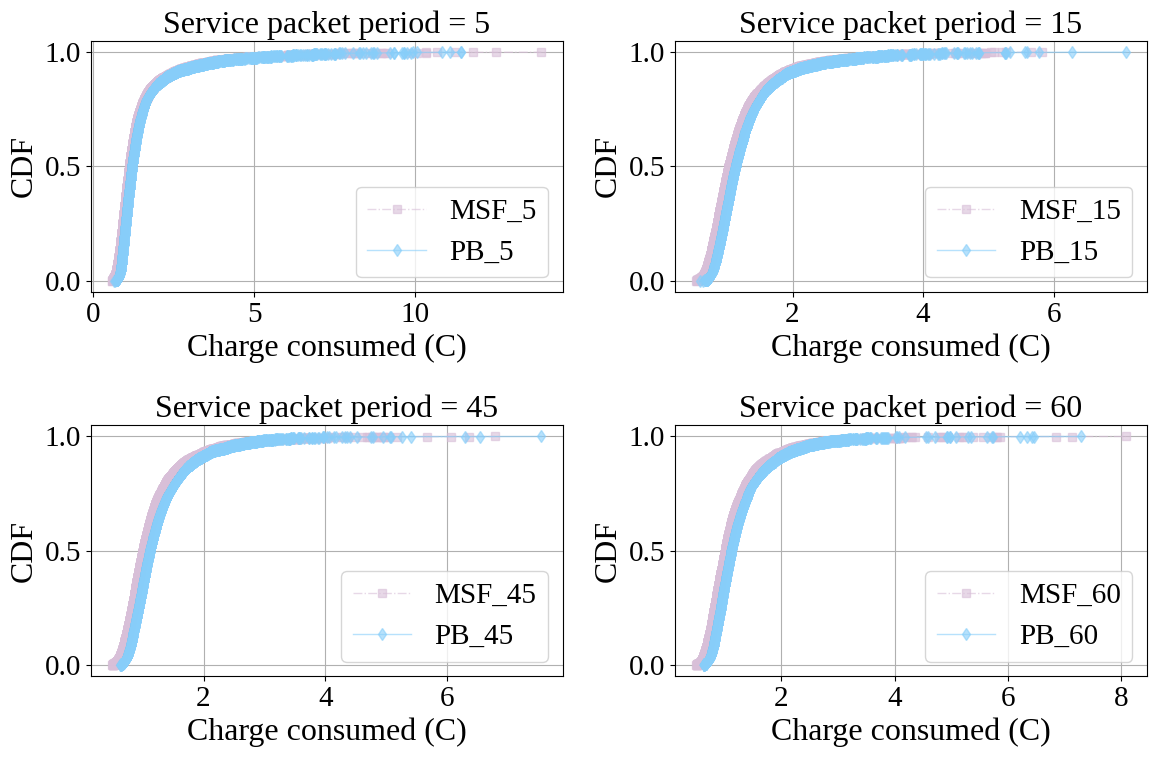}
			\caption{$N = 50$}
			\label{fig:charge_per_node_50nodes}
		\end{subfigure}		
		\begin{subfigure}{0.82\linewidth}
			\includegraphics[width=\textwidth]{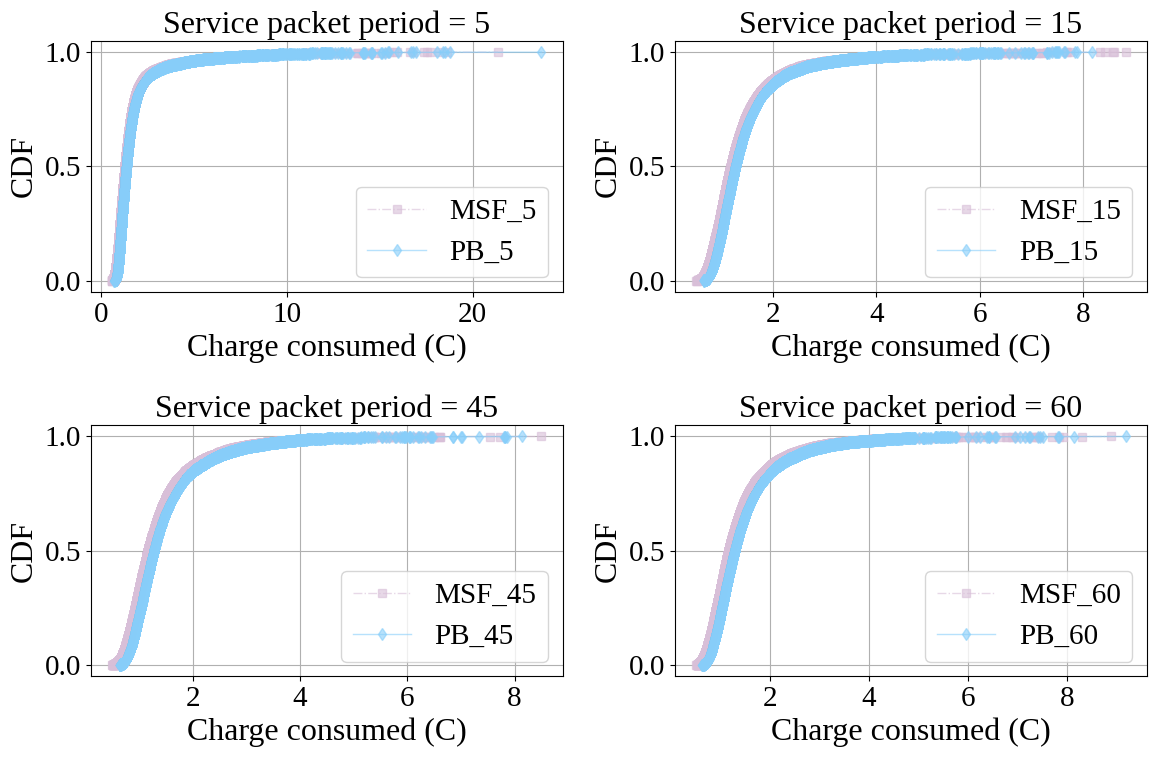}
			\caption{$N = 100$}
			\label{fig:charge_per_node_100nodes}
		\end{subfigure}
		\caption{CDF of consumed charge.}
		\label{fig:charge_node}
	\end{figure}

	\begin{figure}
		\begin{subfigure}{0.82\linewidth}
			\includegraphics[width=\textwidth]{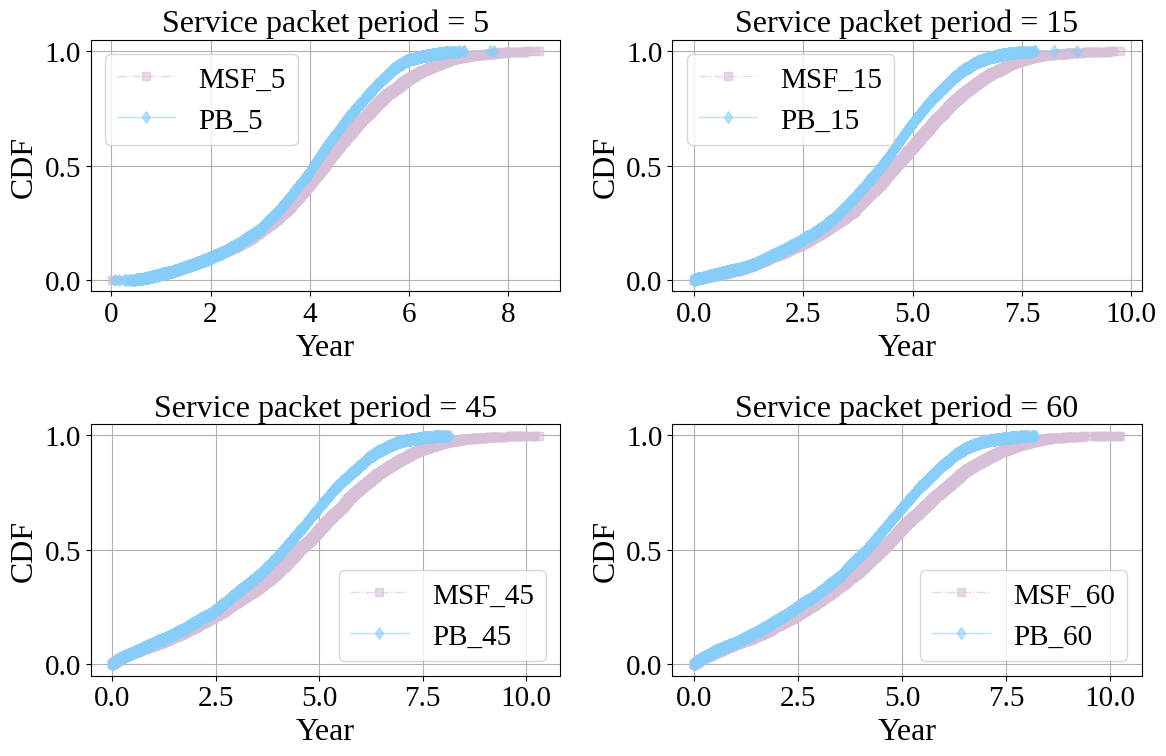}
			\caption{$N = 50$}
			\label{fig:lifetime_50nodes_cdf}
		\end{subfigure}		
		\begin{subfigure}{0.82\linewidth}
			\includegraphics[width=\textwidth]{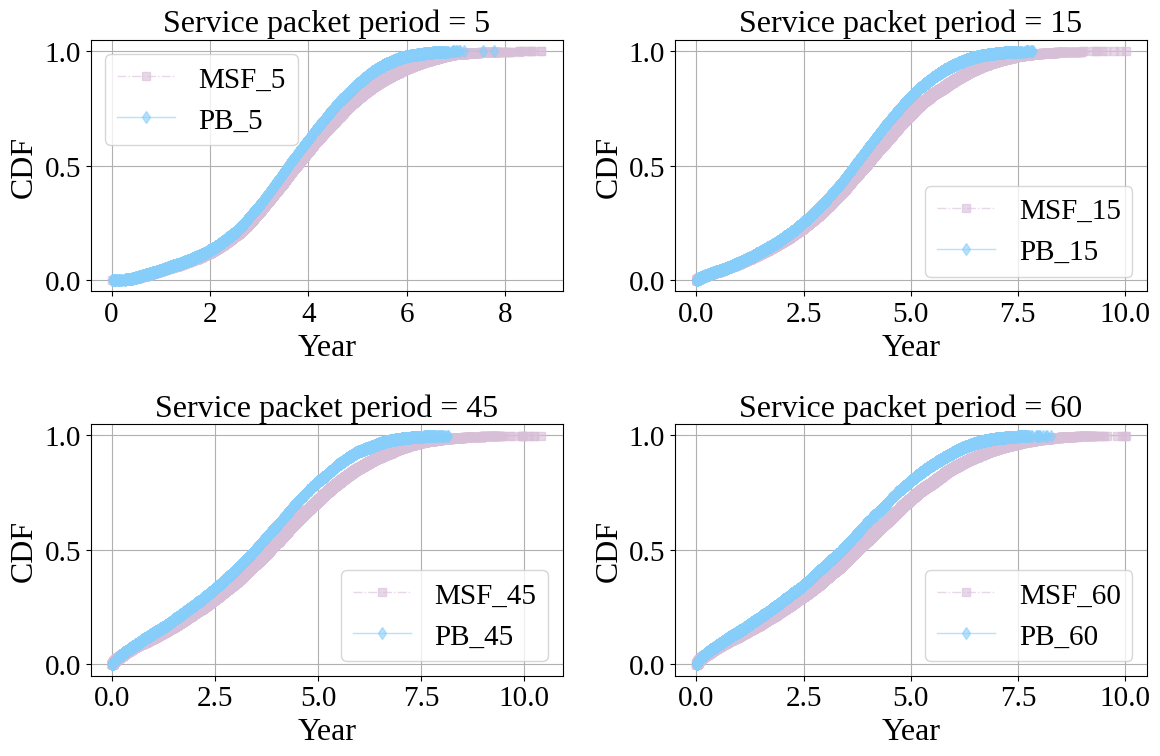}
			\caption{$N = 100$}
			\label{fig:lifetime_100nodes_cdf}
		\end{subfigure}
		\caption{CDF of nodes' lifetime.}
		\label{fig:lifetime_cdf}
	\end{figure}

	{}{Figs. \ref{fig:charge_node} and \ref{fig:lifetime_cdf} present the Cumulative Distribution Function (CDF) of charge consumption and node lifetime, respectively, for PB and MSF under different network sizes ($N=50$, $N=100$) and service packet periods (5, 15, 45, 60s).} {}{The node lifetime is determined by
			\begin{equation}
			\text{Lifetime} = \frac{\text{Battery Capacity (µAh)}}{\text{Average Current Consumption (µA)} \times 24 \times 365},
			\end{equation}
			where the ``\textit{Average current consumption}'' corresponds to the mean current consumption trends presented in Fig. \ref{fig:charge_node}. The CDF in Fig. \ref{fig:charge_node} shows that PB and MSF consume nearly identical charge per node across all scenarios, confirming that PB does not introduce additional energy overhead. However, the CDF in Fig. \ref{fig:lifetime_cdf} reveals that PB nodes exhibit slightly shorter lifetimes than MSF nodes, despite having the same total charge consumption. This suggests that PB nodes experience a higher current consumption per time unit due to increased activity, implying that PB nodes spend less time in low-power states, such as idle listening, sleep, or desynchronization.} {}{For $N=100$, the difference in node lifetime between PB and MSF becomes less pronounced, indicating that PB uniformly distributes energy across nodes in dense networks. The similarity in charge consumption across different network sizes confirms that PB maintains stable energy consumption.} 

{}{Simulation results demonstrate that the proposed solution consistently outperforms the MSF benchmark in terms of end-to-end latency, jitter, number of transmitted/received packets, and energy efficiency across both short-term and long-term scenarios. Notably, our solution maintains stable performance as the network size increases from $N=50$ to $N=100$ nodes, highlighting its scalability for large IIoT networks. The cumulative energy consumption remains consistent with increasing node density, indicating preserved efficiency under varying network conditions. These findings prove our method’s suitability for real-time and energy-constrained applications and its effectiveness in large and dynamic IIoT systems.} 

    \section{Discussion and Future Directions}
    {}{Despite the demonstrated advantages, the proposed approach presents certain trade-offs and limitations.}  {}{For instance, our slot merging strategy is limited by the maximum payload size imposed by TSCH, which must be respected to avoid packet fragmentation. Exceeding this limit leads to fragmentation, thus increasing transmission time, energy consumption, and control overhead while undermining the benefits of packet merging by creating additional fragments that must be transmitted separately. Consequently, fragmentation can cause delays, raise the risk of packet loss, and negate the intended efficiency gains. Moreover, when the size of the injected data grows, the processing time required for slot selection increases. To ensure that slot selection is completed within the TSCH slot duration, it may be necessary to adjust the latter's duration according to the device's computational capability. However, increasing the slot duration reduces the overall network throughput, hence creating a trade-off between scheduling efficiency and transmission capacity. This challenge underscores the need for optimized slot selection algorithms that minimize processing time without extending the TSCH slot duration.}

    {}{Also, in the proposed approach to manage TSCH queue overflow, knowing the reason behind packet accumulation enables more informed decisions, greater efficiency, and increased resilience against edge-case scenarios. Indeed, initiating a reservation request without identifying the underlying cause could worsen the situation. For example, if packet accumulation results from a temporary disconnection with the parent node (disjoin), sending a reservation request would be inefficient and may lead to unnecessary resource consumption. Conversely, if the accumulation is due to a lack of available transmission slots or network congestion, requesting additional cells becomes suitable. Therefore, in our future work, we will focus on developing a decision-making mechanism that diagnoses the cause of packet accumulation to determine the appropriate mitigation strategy.}

    {}{This work serves as a prototype implementation of the proposed approaches, enabling the optimization of key parameters and analysis of their impact on network efficiency. However, transitioning from simulations to real-world deployments introduces additional challenges, such as hardware limitations, environmental interference and obstacles, multi-protocol co-existence, and scalability, which warrant further investigation. In future work, we aim to validate the proposed approach through hardware-based experiments and explore alternative slot encoding methods to tackle time slot duration and TSCH payload constraints, which would enhance the proposed solution's scalability and adaptability in real environments.}
    
    {}{Finally, there is a correlation between the slotframe length, TSCH payload constraints, and time slot duration. This relationship depends on several factors, including the network size and the node’s physical characteristics. In future work, we will investigate this relationship and develop strategies for optimal scheduling, data compression, and adaptive slotframe length configuration to maximize scheduling efficiency.}

    \section{Conclusion}	
	In this paper, we introduced a novel approach that optimizes 
    the performances of the 6TiSCH network. Specifically, through the integration of several proposed changes within the exchanged control messages and operations of SF and RPL, our solution improves {}{network formation by reducing node joining time and energy consumption. Moreover, it reduces end-to-end latency and jitter, while enabling a higher number of successfully received packets compared to the MSF benchmark. Also, it keeps } 
  	 energy usage at levels similar to those of the MSF benchmark. This work provides important insights into the enhancement of 6TiSCH operations for future 5G and beyond IIoT networks. {}{Nevertheless, challenges remain to unleash its full potential, including balancing slotframe length and overcoming the TSCH payload limit. Future work will focus on adaptive scheduling for dynamic slotframe adjustment, intelligent queue management to avoid unnecessary cell reservations, and real-world experimental validation to assess performances under practical conditions.} 

	\section*{Acknowledgment}
	This work was supported by funding from the Algerian Ministry of Higher Education and Scientific Research.
	
	\bibliographystyle{elsarticle-num} 
	\bibliography{6tisch, 6tisch_2022}
	
\end{document}